\documentclass[fleqn,usenatbib]{mnras}

\usepackage{newtxtext,newtxmath}

\usepackage[T1]{fontenc}

\DeclareRobustCommand{\VAN}[3]{#2}
\let\VANthebibliography\thebibliography
\def\thebibliography{\DeclareRobustCommand{\VAN}[3]{##3}\VANthebibliography}

\usepackage{graphicx}	
\usepackage{amsmath}	
\usepackage{hyperref}
\usepackage{color}
\usepackage{longtable}

\title[Characterising HOD in filaments and nodes of the cosmic web]{Characterising HOD in filaments and nodes of the cosmic web}

\author[Perez et al.]{
Noelia R. Perez,$^{1}$\thanks{E-mail: noeliarocioperez@gmail.com (NP)}
Luis A. Pereyra,$^{2,3}$
Georgina Coldwell,$^{1}$
 Facundo Rodriguez,$^{2,3}$
\newauthor Ignacio G. Alfaro$^{2,3}$
and  Andr\'{e}s N. Ruiz$^{2,3}$
\\
$^{1}$Facultad de Ciencias Exactas, F\'{i}sicas y Naturales, Departamento de Geof\'{i}sica y Astronom\'{i}a, CONICET Universidad Nacional de San Juan, Av. Ignacio de la\\  Roza 590 (O), J5402DCS, Rivadavia, San Juan, Argentina\\
$^{2}$Instituto de Astronom\'{i}a Te\'{o}rica y Experimental, CONICET-UNC, Laprida 854, X5000BGR C\'{o}rdoba, Argentina\\
$^{3}$Observatorio Astron\'{o}mico de C\'{o}rdoba, UNC, Laprida 854, X5000BGR C\'{o}rdoba, Argentina
}

\date{Accepted XXX. Received YYY; in original form ZZZ}

\pubyear{2015}

\begin{document}
\label{firstpage}
\pagerange{\pageref{firstpage}--\pageref{lastpage}}
\maketitle

\begin{abstract}

The standard paradigm for the formation of the Universe suggests that large structures are formed from hierarchical clustering by the continuous accretion of less massive galaxy systems through filaments.
In this context, filamentary structures play an important role in the properties and evolution of galaxies by connecting high-density regions, such as nodes, and being surrounded by low-density regions, such as cosmic voids. The availability of the filament and critical point catalogues extracted by \textsc{DisPerSE} from the \textsc{Illustris} TNG300-1 hydrodynamic simulation allows a detailed analysis of these structures. The halo occupation distribution (HOD) is a powerful tool for linking galaxies and dark matter halos, allowing constrained models of galaxy formation and evolution.
In this work we combine the advantage of halo occupancy with information from the filament network to analyse the HOD in filaments and nodes.
In our study, we distinguish the inner regions of cosmic filaments and nodes from their surroundings.
The results show that the filamentary structures have a similar trend to the total galaxy sample covering a wide range of densities.
In the case of the nodes sample, an excess of faint and blue galaxies is found for the low-mass halos suggesting that these structures are not virialised and that galaxies may be continuously falling through the filaments. 
Instead, the higher-mass halos could be in a more advanced stage of evolution showing features of virialised structures. 

\end{abstract}




\section{Introduction}
In the current Standard Model of Cosmology, $\Lambda$ Cold Dark Matter ($\Lambda$CDM) baryons and dark matter accounts roughly $30\%$ of the total energy-matter, which are arranged in a complex and massive network known as the ``Cosmic Web''
\citep{deLapparent1986,bond96}.
The web pattern is composed of nodes, filaments and voids, and its study allows us to understand the evolutionary process of the Universe. 

Filaments are traced by galaxies and transport matter 
from voids and walls to nodes \citep{Zel'dovich1970,Cautun2014}.
Nodes are typically found in regions of high density and voids inhabit regions of significantly lower density. 
As for to filaments, they span a wide range of densities, from superdense regions (though lower than nodes) to subdense regions (though higher than voids). 
The density threshold is therefore not sufficient to characterise a particular cosmic environment \citep{Cautun2014}.

In this context, 
\cite{shandarin89} suggest that dark matter halos are formed from initial density perturbations in the density field of the early Universe and grow with time due to gravitational instability \citep{Lifshitz46,Peebles1980}. 
Later, when these perturbations are dissociated from the expansion, they collapse into dark matter halos, which continue to accrete material through hierarchical clustering by continuously merging smaller structures. \citep{WhiteRees78,White1994}.

The galaxy formation and evolution are influenced by the environment, so it is expected that the large-scale surrounding affects their properties.
Several works reveal the impact of these structures on galaxy shape and spin alignment \citep{Faltenbacher2002,Trujillo2006,Aragon-Calvo_2007,Tempel2013,Tempel2013_2,Forero-Romero2014,Zhang2015,Ganeshaiah_Veena2018,GaneshaiahVeena2019,Wang2020,Lee2023}, and on properties such as galaxy stellar masses, star formation rate (SFR) and colour \citep{Weinmann2006,Einasto2008,Lietzen2012,Darvish2016,Malavasi2017,Kraljic2018,Laigle2018}.

The formation and evolution of galaxies in dark matter halos depend on a variety of physical processes, and also because their properties are related to the host halo features. In this respect, it is not possible to determine exactly how galaxies populate the dark matter halos in which they reside.
The Halo Occupation Distribution (HOD) model empirically links galaxies and dark matter halos, characterising the probability distribution $P(N|M)$ that a virialized halo of mass $M$ contains $N$ galaxies of previously determined characteristics \citep{Peacock2000,Berlind2002,Berlind2003,Zheng2005,Guo2015,Rodriguez2015, Rodriguez2020}.
Moreover, \cite{Peacock2000} argued that the distribution of galaxies within the halo density field depends on their position and the number of objects within the host halo. 
In this sense, the HOD formulation assumes that the host halo mass is the leading property that influences the galaxy properties \citep{WhiteRees78}, although the spatial distribution of galaxies also depends on other properties (formation time, spin, concentration, neighbour mass, large-scale structure), known as the assembly bias \citep{Gao2005,Gao2007,Mao2018,Salcedo2018,Musso2018,Ramakrishnan2019,Mansfield2020}.

The HOD framework has been useful in constraining models of galaxy formation and evolution \citep{Berlind2003,Kravtsov2004,Zehavi2018}, and cosmological models \citep{vandenBosch2003,Zheng2007} throughout the study of the dependence of the galaxy population properties \citep{Berlind2003,Zheng2005,Contreras2013,Borzyszkowski2017,Zehavi2018,Yuan2022}, and clustering variations with respect to large-scale environments \citep{Artale2018,Zehavi2018,Alfaro2020, Alfaro2021,Alfaro2022}. Furthermore, by characterising the evolution of the clustering
\citep{Contreras2023}, interpreting the clustering data \citep{Kim2009,Ross2010,Zehavi2011} and creating mock catalogues \citep{Grieb2016} the halo population can be better understood.


We stress that our aim in this paper is to identify any differences that may exist for HODs
in different cosmic web structures, taking into account a wide range of magnitude thresholds. 
We are also exploring the HOD considering different galaxy properties such as colour 
and morphology, in order to provide a comprehensive characterisation. 
This paper is structured as follows: in Section \ref{sec:data} we summarise the data and describe the criteria for selecting samples. 
In Section \ref{sec:analysis} we analyse the  HOD variations in nodes and filamentary structures, and their respective outskirts considering different magnitude limits. 
In this section, we also explore and compare the HOD for these environments with their surroundings taking into account diverse galaxy properties. 
Finally, in Section \ref{sec:conclusions} we summarise and discuss the main results.


\section{Data} \label{sec:data}
\subsection{IllustrisTNG}
This analysis has been based on the \textsc{IllustrisTNG} project\footnote{https://www.tng-project.org/} \citep{Marinacci2018,Naiman2018,Nelson2018,Pillepich2018_a,Pillepich2018_b,Springel2018}, a suite of cosmological hydrodynamic simulations developed from the original \textsc{Illustris} simulations \citep{Genel2014,Vogelsberger2014_a,Vogelsberger2014_b} and executed with the \textsc{arepo} moving mesh code \citep{Springel2010}. 
The cosmological parameters are in agreement with Planck 2015 results  
\citep{planckresults}: $\Omega_{\text{m},0}=0.3089$, $\Omega_{\Lambda,0}= 0.6911$, $\Omega_{\text{b},0}= 0.0486$, $\text{h}= 0.6774$, $\text{n}_{\text{s}}= 0.9667$ and $\sigma_{8}= 0.8159$.
The project has three sizes of physical simulation boxes, each with a side length of approximately $50$, $100$ and $300\ \text{Mpc}$, named TNG50, TNG100 and TNG300, respectively.

In particular, we used TNG300-1 because its higher resolution makes it the most suitable simulation for studying large-scale structures.
TNG300-1 employs $2500^3$ dark matter particles and $2500^3$ gas particles with masses of $5.9 \times 10^7 \text{M}_{\odot}$ and $1.1 \times 10^7 \text{M}_{\odot}$, respectively, in a box of $205\ h^{-1} \text{Mpc}$.
There are 100 snapshots available and each one is associated with a group catalogue containing halos and subhalos.
Halos (also named FoF Halo, FoF Group or Group) are found using the friend-of-friend algorithm \citep{Huchra1982} with linking length $b = 0.2$ and the \textsc{subfind} algorithm \citep{Springel2001} is used to identify subhalos (also known as Subfind Group, Subgroup or Galaxies)

\subsection{DisPerSE}

The Discrete Persistent Structures Extractor, \textsc{DisPerSE} \footnote{http://www2.iap.fr/users/sousbie/web/html/index888d.html?archive} \citep{Sousbie2011_a,Sousbie2011_b,Sousbie2013} is a multiscale identifier of structures based on discrete Morse theory.
According to this theory, the Cosmic Web can be characterised in a mathematical equivalent called the Morse complex.

Initially, the algorithm generates the density field of a discrete set of points through the Delaunay Tessellation Field Estimator (DTFE, \citet{Schaap2000,vandeWeygaert2009}). This allows the identification of critical points, i.e. points where the density gradient is zero (maxima, minima and saddle points).
Then, Morse's theory uses mathematical properties that describe the relationship between topology and the geometry of the density field. The critical points together with their ascending and descending manifolds $0$, $1$, $2$ and $3$ can be related to clusters, filaments, walls and voids, respectively. 
To quantify the importance of the topological features found, the user can select the persistence level for filtering the Poisson sample noise and intrinsic uncertainty within the data set. 
%

For this work, we use the 
filaments and critical point catalogues developed by \cite{Duckworth2020a,Duckworth2020b}
\footnote{https://github.com/illustristng/disperse\_TNG} 
using the \textsc{DisPerSE} code, and available for TNG300-1.
To build these catalogues, the authors selected subhalos with masses higher than $\text{M}_* > 10^{8.5}\ h^{-1} \text{M}_{\odot}$ and set the persistence parameter $\sigma = 4$ to remove spurious filaments. The catalogues are available for 8 snapshots between $z=0$ and $z=2$ and provide the cosmic web distances between each subhalo to the nearest minimum critical point, the nearest 1-saddle point, the nearest 2-saddle point, the nearest maximum critical point and the nearest filament segment \citep[see for instance][]{Duckworth2020a}.

\subsection{Sample selection}

The sample used in this paper was constructed from the TNG300-1 Groupcat at $z=0$ by selecting subhalos with $\text{M}_* > 10^{8.5}\ h^{-1} \text{M}_{\odot}$, in agreement with the filament catalogue
, contained in halos with masses $\text{M}_{200} \geq 10^{11}\ h^{-1} \text{M}_{\odot}$, where ${\rm M}_{200}$ is the mass enclosed within a region that encompasses $200$ times the critical density. The purpose of this selection is to obtain well-resolved halos with about $10^3$ particles. 
The final catalogue contained 212749 halos and 264407 subhalos.

With the aim of determining any existing differences in the halo occupation with respect to the a variety of cosmic environments, we have separated the halos according to whether they belong to regions such as nodes and filaments, respectively. 
%

In the following, we will refer to nodes as halos with a distance to maxima (peaks of the DisPerSE density field) less than $\rm R_{\rm 200}$  ($\text{d}_{\text{n}} \leq 1 \text{R}_{\text{200}}$), where $\rm R_{\rm 200}$ is the comoving radius of a sphere centred on the halo whose mean density is $200$ times the critical density of the Universe. Meanwhile, for the filaments sample, we selected halos with distance to the node $\text{d}_{\text{n}} > 1 \text{R}_{\text{200}}$ and distance to the filament axis $\ \text{d}_{\text{f}} \leq 1 h^{-1} \text{Mpc}$.
This value was chosen to ensure that the objects belong to the filament core \citep{GalarragaEspinosa2022,GalarragaEspinosa2023}.

To associate each halo and subhalo with a certain environment, we determined the distances from these objects to the nearest node and filament axis. 
Table \ref{tab:num_halos} summarises the percentage of halos and subhalos for each defined sample. The outskirt sample definitions are described in sections \ref{ssec:Filaments} and \ref{ssec:Nodes}. 
In addition, we show the percentages of halos and subhalos not selected as part of our definitions in this work (Unclassified).
We can observe that approximately half of the halos are found in filament environments (filaments and filament outskirts), less than $10\%$ of the halos inhabit node regions (nodes and node outskirts) and the rest are found in structures not considered in this work. 
This result is in agreement with \cite{Cautun2014,GaneshaiahVeena2019}, who found that the dark matter fraction is contained in filaments, walls, voids and nodes with 50, 25, 15 and 10 per cent, respectively.
Although in the mentioned works, the fractions are calculated using the particles and not from the halos.


\begin{table}
	\centering
	\caption{Percentage of halos and subhalos in each sample.}
	\label{tab:num_halos}
    \setlength{\tabcolsep}{4pt} 
	\begin{tabular}{lccccc}
        \hline
  		   & Filaments & Nodes & Filament & Node & Unclassified\\
          &       &           & outskirts & outskirts & \\
          & (per cent)&(per cent) & (per cent) & (per cent) & (per cent)\\

		\hline
        Halos    & 29.57 &  3.31 & 20.46 & 3.79 & 42.87 \\
		Subhalos & 29.23 & 30.71 & 13.32 & 2.81 & 23.93 \\
		\hline
	\end{tabular}
\end{table}

\section{Analysis} \label{sec:analysis}

\begin{figure*}
\centering
    {
        \includegraphics[width=0.47\textwidth]{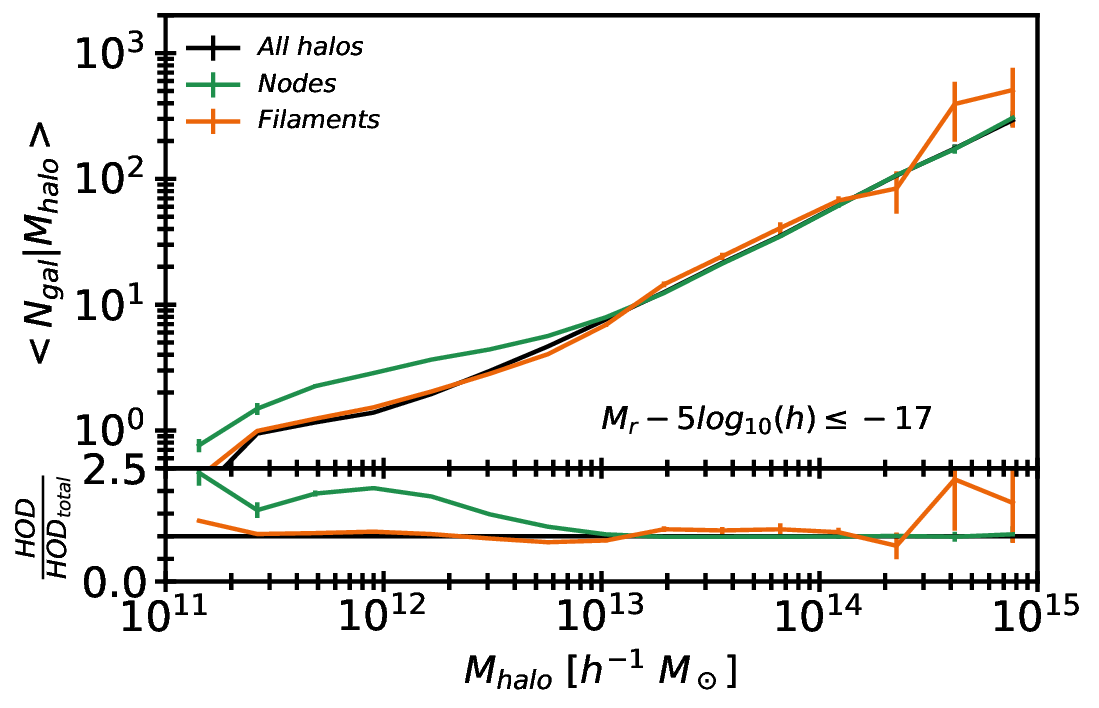}
    }
    {
        \includegraphics[width=0.47\textwidth]{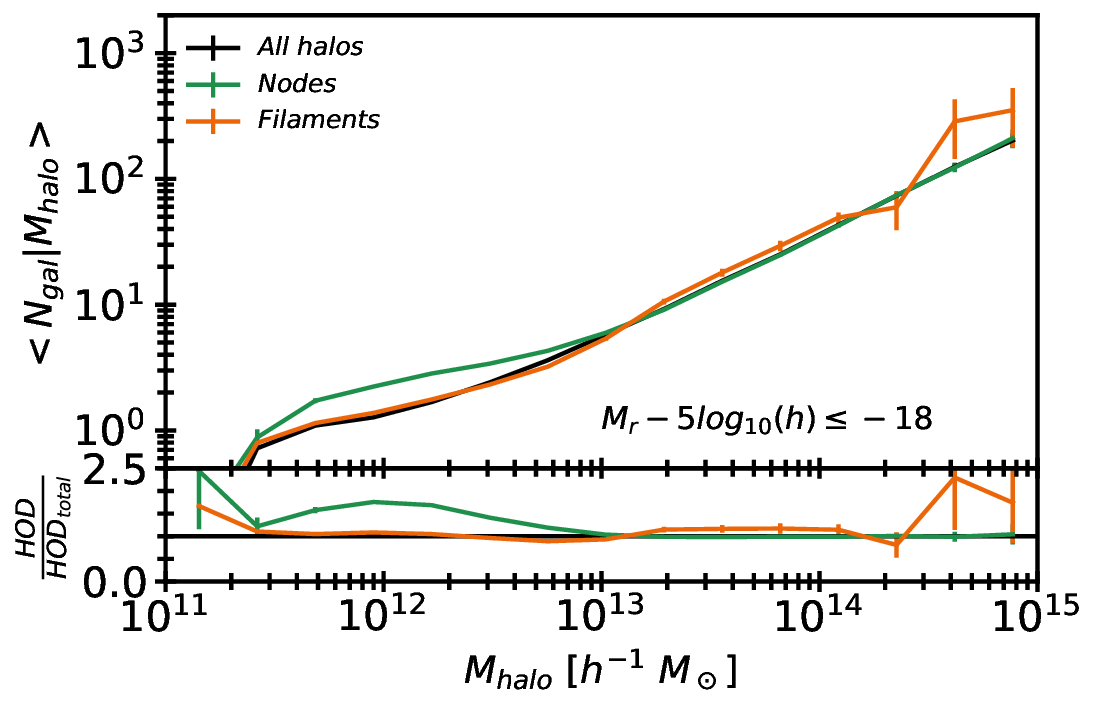}
    }
    {
        \includegraphics[width=0.47\textwidth]{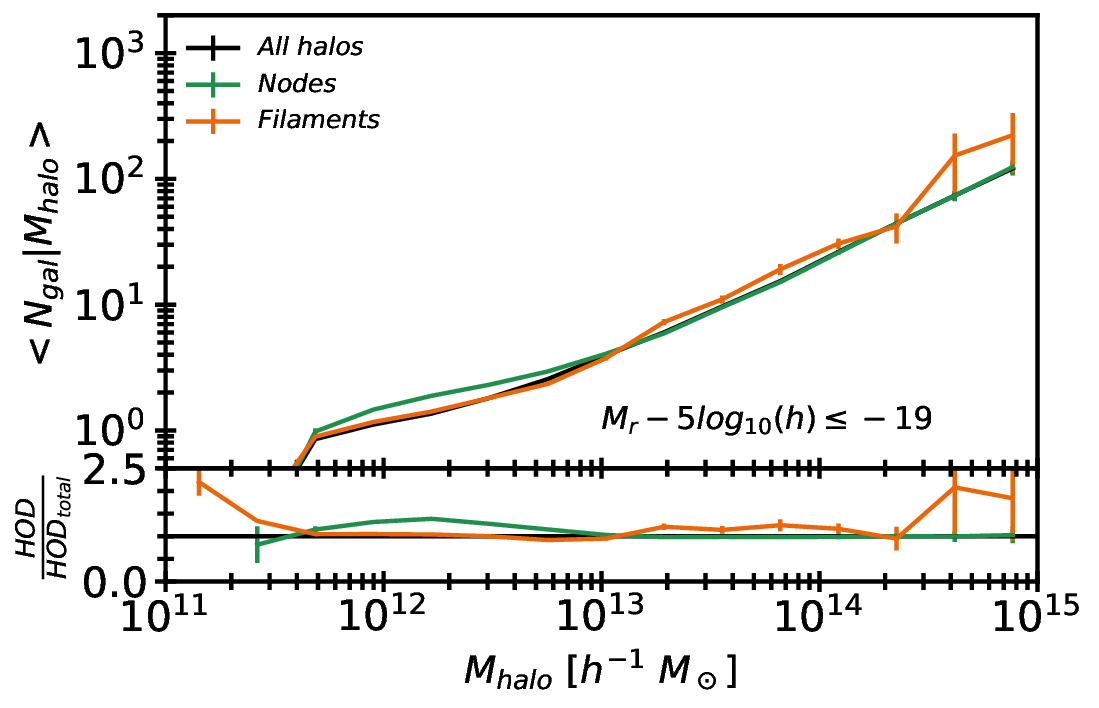}
    }
    {
        \includegraphics[width=0.47\textwidth]{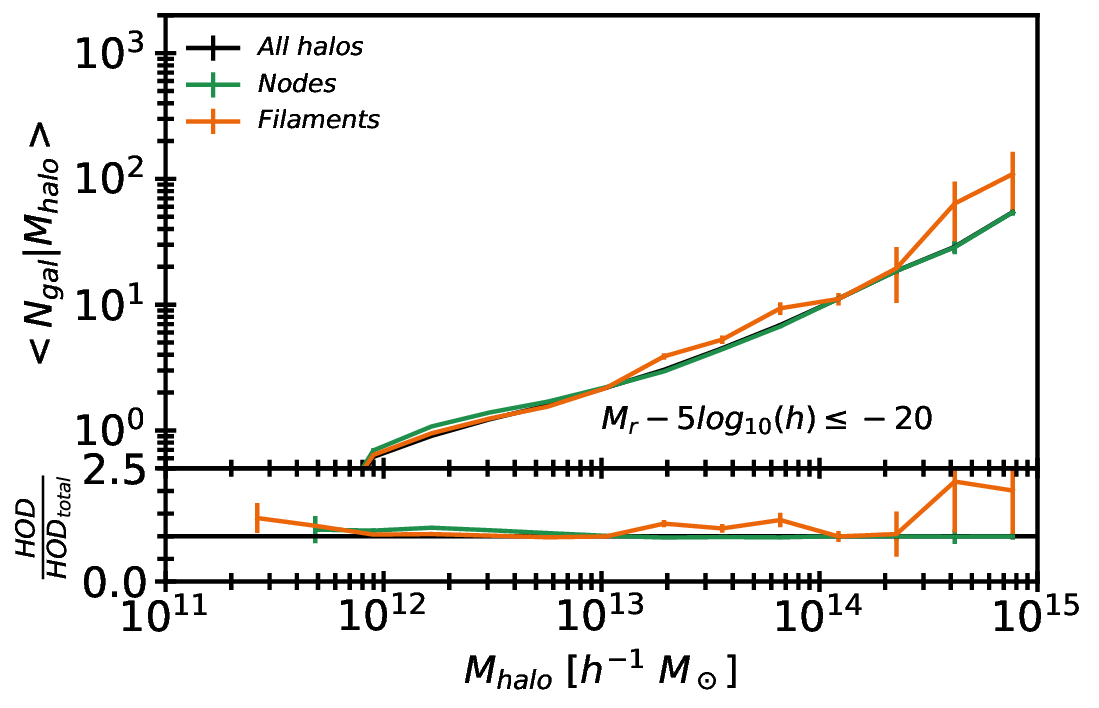}
    }
    \caption{The panels show the HODs measured for galaxies with different magnitude thresholds for the whole sample (black lines) and for galaxies belonging to filaments (orange lines) and nodes (green lines), respectively. Lower sub-panels show the ratio between the HOD inside filaments and nodes and the HOD in the whole sample.}
    \label{fig:hod_general}
\end{figure*}

\begin{figure*}
    \centering
    \includegraphics[width=0.7\textwidth]{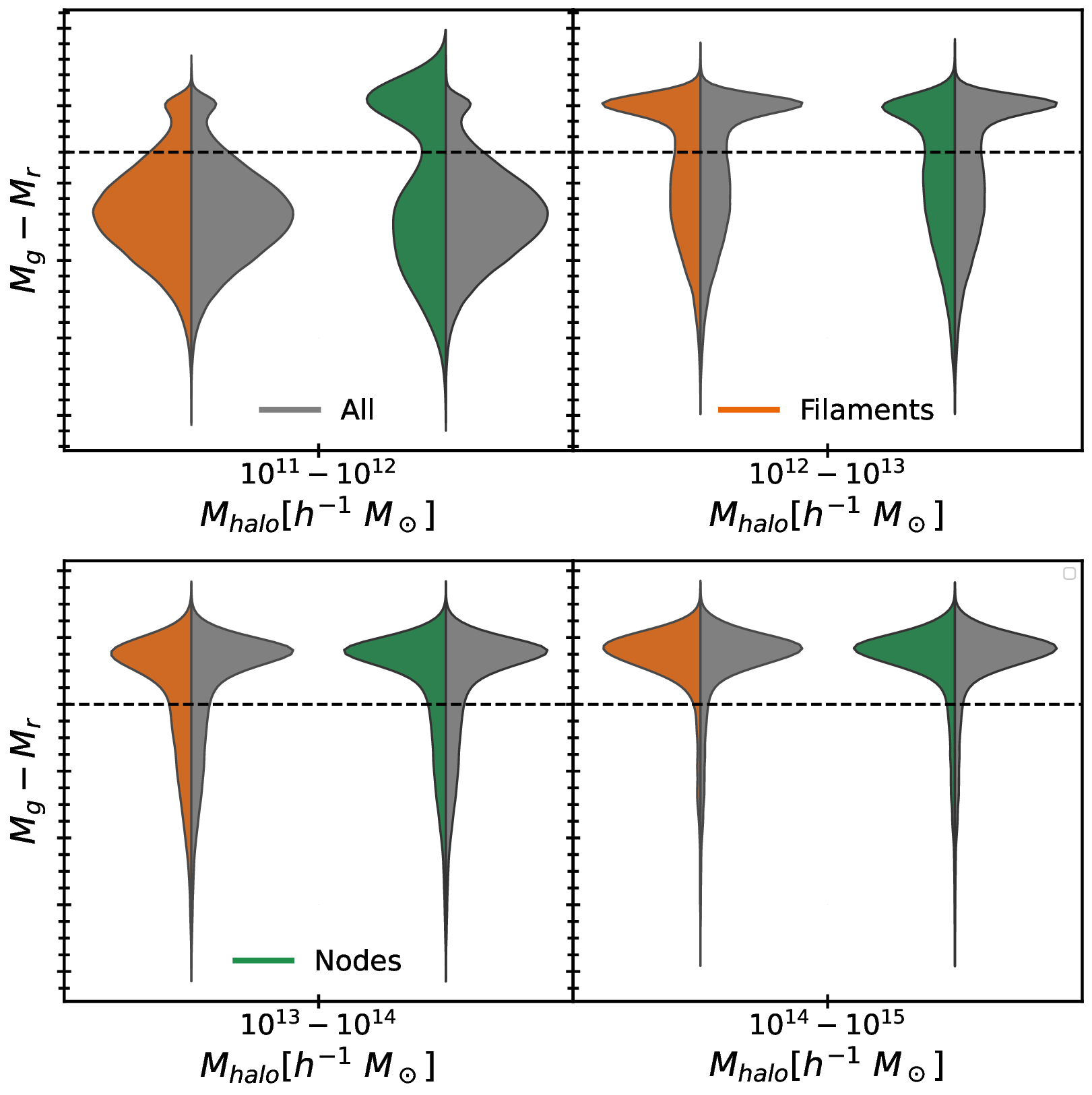}    
    \caption{Galaxy colour distributions, in the halo mass bins, for the total sample (grey), the filaments sample (orange) and the nodes sample (green).}
    \label{fig:distr_gr_masas}
\end{figure*}

\begin{figure}
\centering
    {
        \includegraphics[width=0.47\textwidth]{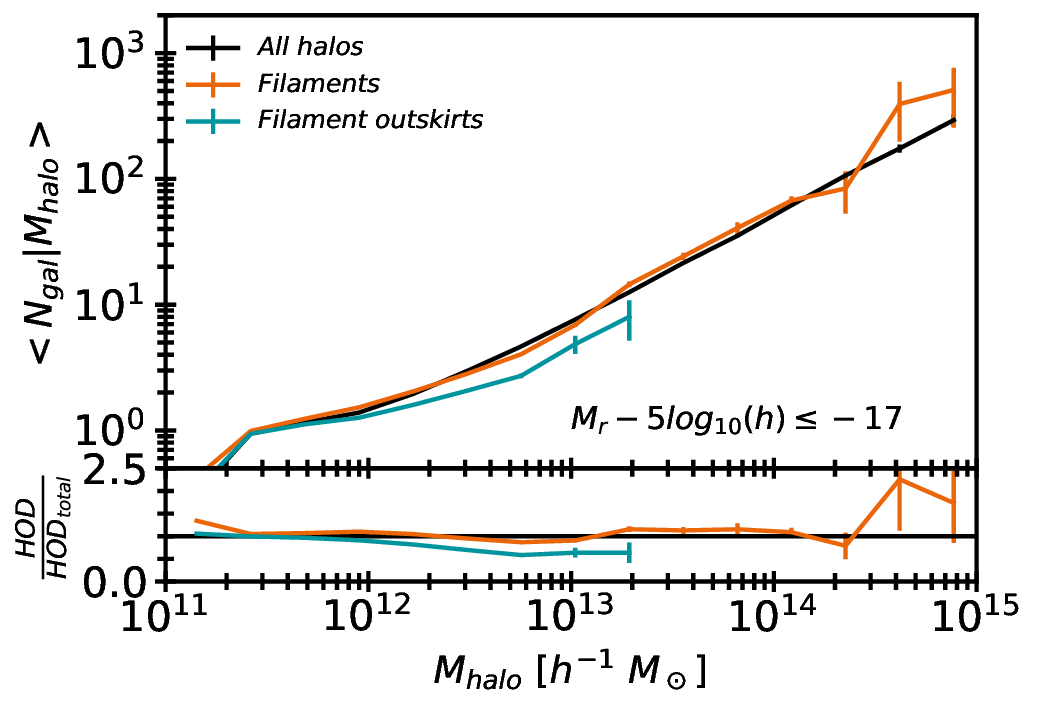}
    }
    {
        \includegraphics[width=0.47\textwidth]{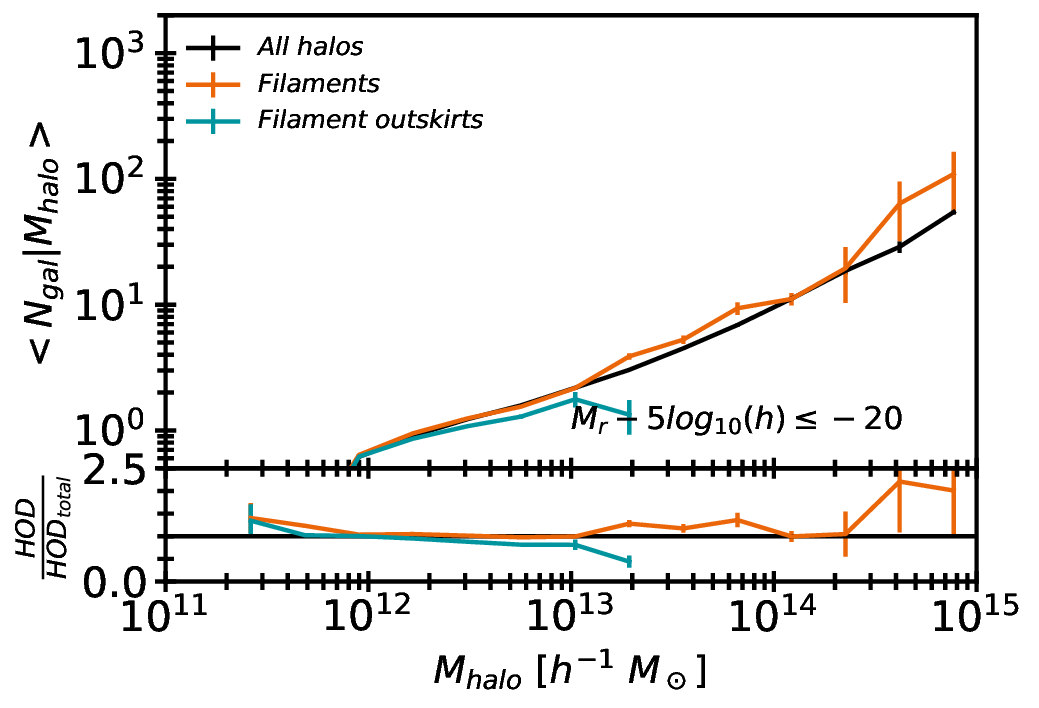}
    }    
    \caption{HODs measured for the total (black), filaments (orange) and filament outskirts (light-blue) samples for subhalos with $\text{M}_\text{r} - 5\text{log}_{10}(\text{h}) \leq -17$ (top panel) and $\text{M}_\text{r} - 5\text{log}_{10}(\text{h}) \leq -20$ (bottom panel), respectively. Sub-panels: the ratio between the filaments and filament outskirts samples respect to the overall HOD.}
    \label{fig:hod_fil}
\end{figure}

\begin{figure*}
\centering
    {
        \includegraphics[width=0.8\textwidth]{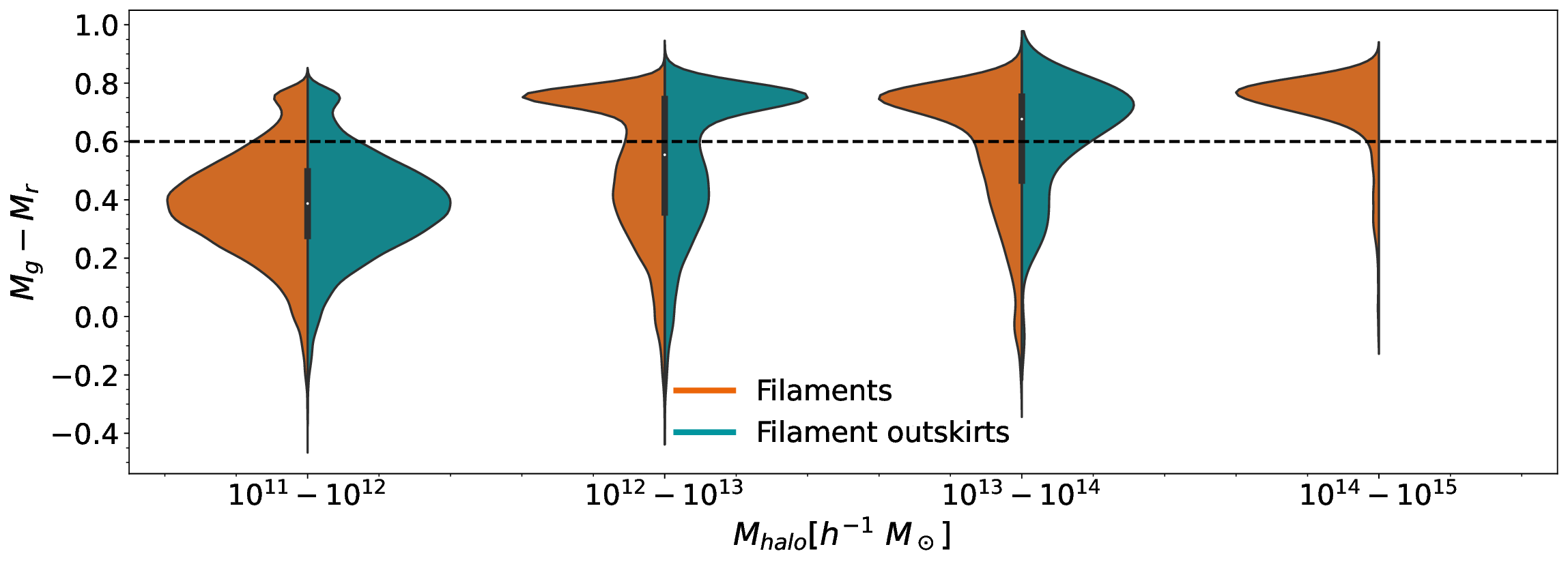}
    }
    \caption{Colour distributions of galaxies in halo mass bins for filaments (orange) and filament outskirts (light-blue) samples.}
    \label{fig:distr_gr_masas_fil}
\end{figure*}

The HOD allows us to statistically determine how many galaxies inhabit halos with a mass within a given range \citep{Berlind2002}.
In this line, \cite{Alfaro2020} studied the HOD for central and satellite galaxies in very low-density environments, given by cosmic voids. They found a significant dependence of the halo occupation behaviour with the mass and luminosity of the galaxies. 
In addition, \cite{Alfaro2021} studied the HOD in a set of future virialised superstructures \citep[FVS,][]{Luparello2011} identified on a cosmological semi-analytic simulation to analyse the occupation of galaxies in halos within globally high-density regions. The authors found that the HOD from high-mass halos significantly increases towards the central regions of these superstructures.

In this section, we describe the procedure for the measurement of the HOD in different cosmic web structures, from the samples selected from the catalogues of the filament and point critics \citep{Duckworth2020a,Duckworth2020b} as described in the previous section, and explore its dependence on several galaxy properties.
To estimate the HOD, we compute the mean number of galaxies per halo mass bins, $\left< \text{N}_{\text{gal}} | \text{M}_{\text{halo}} \right>$, considering different ranges of $r$-band absolute magnitude for the subhalo. 
The error bars were determined using the jackknife technique \citep{Quenouille1949,Tukey1958}.

Following \cite{Alfaro2020}, we use different absolute magnitude limits to compare the behaviour of faint and bright galaxies in the HOD. We selected the $r$-band because it is an estimator of stellar mass and could be compare with the observational data. The magnitude values used correspond to SubhaloStellarPhotometrics catalogue \citep{Nelson2018}.
In Fig. \ref{fig:hod_general}, the HODs are shown for the total, nodes and filaments samples considering the following $r$-magnitude thresholds: $\text{M}_\text{r} - 5\text{log}_{10}(h) \leq -17, -18, -19\ \text{and} -20$.
The lower sub-panels show the quotient between the nodes and filaments samples with respect to total HOD. 
As expected, the halo occupancy decreases when we consider increasingly brighter subhalos.
The HOD for halos belonging to filaments is quite similar to that shown for the total sample, independent of galaxy luminosity.
An increase in error can be observed for halos with masses higher than $10^{14.5}\ h^{-1} \text{M}_{\odot}$.

On the other hand, for the nodes sample, there is a significant excess of faint galaxies compared to the filaments and total samples.
The main difference in the HOD between the samples occurs for halos with masses lower than $10^{13}\ h^{-1} \text{M}_{\odot}$, where we see that the HOD measured for the nodes is approximately two times higher than that measured for the filaments for the magnitude limit: $\text{M}_\text{r} - 5\text{log}_{10}(h) \leq -17$.
This discrepancy decreases and becomes imperceptible as we look at brighter galaxies, so this dependence on the magnitude thresholds could be related to the presence of satellite galaxies in the lower-mass halos.
The distributions for halos with masses $\text{M}_{\text{halo}} > 10^{13}\ h^{-1} \text{M}_{\odot}$ are comparable.
The error bars in the extreme regions, especially for the filaments sample, is related to the fact that there are few massive halos in these structures, so the last bins are almost free of objects.

The study conducted by \cite{Alfaro2021} in high-density regions (FVS) reveals a different trend compared to the nodes examined in this work. In particular, the FVS showed an increase in the mean number of galaxies in halos with masses greater than $\sim 10^{13}h^{-1}M_{\odot}$, regardless of their luminosity, whereas no significant deviations were observed in the nodes compared to the whole halo sample.
On the contrary, for halo masses smaller than about $10^{13}h^{-1}M_{\odot}$,  the HOD within the FVS shows negligible fluctuations. 
Instead, the HODs of nodes demonstrate significant differences in low-mass haloes, with a stronger luminosity dependence for the magnitude threshold of $\text{M}_\text{r} - 5\text{log}_{10}(h) \leq -17$. Then, to enable a more comprehensive analysis within this luminosity range, we have set this limit, and also the limit corresponding to the brightest $r$-magnitude threshold, for analysis in the upcoming sections.
%
%

Since the differences detected in the HODs could be related to the properties of the galaxies, we performed an exhaustive analysis using galaxy colours since this property provides us with information on morphology, age and evolution \citep{Blanton2005,Vulcani2015}.
In Fig. \ref{fig:distr_gr_masas}, we plotted the galaxy colour distributions for filaments, nodes and total samples considering the halo mass bins.
In accordance with \cite{Nelson2018},
the general tendency suggests that low-mass halos are mainly populated by blue galaxies, with a small contribution from the red ones. 
Furthermore, as the halos become more massive, the number of red galaxies increases while the number of blue galaxies gradually decreases. Finally, massive halos are populated by red galaxies, suggesting the presence of a more evolved population.
Regarding the nodes and filaments samples,  
for massive halos, the galaxy colours in both samples are comparable to the total sample.
As for the low-mass halos, we found that the filaments sample follows the overall trend, whereas the nodes sample shows a remarkable bimodal colour distribution. 
This behaviour could be explained by the fact that the low-mass nodes represent local density maxima in the cosmic web inhabited by galaxies  groups, which show evidence of physical processes that redden them even before they reach the virial radius \citep{Blanton2003,Kraljic2018}.

%
To complement the analysis, 
we determine the morphology using the supplementary catalogue of stellar circularities, angular momentum, and axis ratios \citep{Genel2015}. 
The morphology, together with colour and star-formation rate (SFR) allows us to characterise the galaxy properties within specific halos.
The correlation between these parameters indicates that late blue galaxies have a higher SFR than early red galaxies (e.g. \citet{Schawinski2014}).

In order to calculate the HODs, we separate the subhalos with respect to the parameters mentioned above.
To exploit the bimodality of galaxy colours, we choose a value of $\text{M}_{\text{g}} - \text{M}_{\text{r}}=0.6$ as the cut-off to distinguishing red ($\text{M}_{\text{g}} - \text{M}_{\text{r}} > 0.6$) from blue ($\text{M}_{\text{g}} - \text{M}_{\text{r}} \leq 0.6$) galaxies \citep{Nelson2018}.
Then, to define disky (spheroidal) galaxies, we use the circularity parameter as a criterion, defining the bulge mass fraction as twice the stellar mass fraction with a negative circularity parameter lower (higher) than $0.5$ \citep[see e.g.][]{Osato2023}.

In the next subsections, we analyse the HODs for filaments, nodes and their surrounding regions separately, in order to study both environments in more detail.
For completeness, the Appendix presents the HODs
acquired using the SFR parameter to distinguish between
quiescent and star-forming galaxies.

\subsection{Filaments}\label{ssec:Filaments}

Taking into account the previous results, in this subsection we have focused on the analysis of the HOD in filamentary structures, considering the variations with respect to the distance to the filament axis.
Then, we consider the filaments sample, which remains unchanged, and the filament outskirts sample defined by selecting halos with $\text{d}_{\text{n}} > 1\ \text{R}_{\text{200}}$ to exclude the nodes and $1  < \text{d}_{\text{f}}\ [h^{-1} \text{Mpc}] \leq 2 $, considering an intermediate region between filaments and voids.
The number of halos and subhalos in each sample is given in Table \ref{tab:num_halos}.

The HODs for filaments and filament outskirts are shown in Fig. \ref{fig:hod_fil}.
In upper panel we show the HOD for galaxies with $\text{M}_\text{r} - 5\text{log}_{10}(h) \leq -17$.
It can be observed that for halo masses smaller than $10^{12}\ h^{-1} \text{M}_{\odot}$ the tendency of both samples is similar following the overall trend. This fact suggests that the process by which galaxies populate these halos may be similar, regardless of the environment.
For $\text{M}_{\text{halo}} > 10^{12}\ h^{-1} \text{M}_{\odot}$, the filaments and total HODs are indistinguishable within the error bars, while for the filament outskirts sample the curve is lower and corresponds to an environment without the presence of massive halos.
This result is in agreement with the trend found by \cite{Alfaro2020}, who studied the HOD in environments with $10$ per cent of the mean density of large scale tracers, reaching a halo mass of $\sim 10^{13}\ h^{-1} \text{M}_{\odot}$, a similar mass range to the filament outskirts. Furthermore, they found that for halos more massive than $10^{12}\ h^{-1} \text{M}_{\odot}$, the HOD in voids it is up to $50\%$ lower than the overall trend.
The lower panel of Fig. \ref{fig:hod_fil} shows the HOD taking into account only the brighter galaxies, i.e. galaxies with $\text{M}_\text{r} - 5\text{log}_{10}(h) \leq -20$. For halos with $\text{M}_{\text{halo}} < 10^{13}\ h^{-1} \text{M}_{\odot}$, the distributions of both samples overlaps the HOD corresponding to the total sample. 
Then, in the outskirts, the HOD shows a decrease, which is related to the small number of halos in the mass range of $10^{13} < \text{M}_{\text{halo}} [h^{-1} \text{M}_{\odot}] < 10^{14}$. Respect to the filaments, the HOD covers the entire mass range and varies within the error bars with respect to the total sample.
We found no significant differences when considering faint and bright galaxies separately. 
That is, the HODs for both filamentary environments show similar behaviour with respect to the magnitude of the galaxies considered.
Then, for the subsequent analysis, we will investigate galaxies with $\text{M}_\text{r} - 5\text{log}_{10}(h) \leq -17$, encompassing a more comprehensive range of magnitudes.

On the other hand,  we explore the galaxy colour distributions, with respect to halo mass bins, for filaments and their outskirts samples. In Fig. \ref{fig:distr_gr_masas_fil}, it can be seen that both environments have comparable distributions for $\text{M}_{\text{halo}} < 10^{13}\ h^{-1} \text{M}_{\odot}$. 
Also, for masses in the range $10^{13} < \text{M}_{\text{halo}}\ [h^{-1} \text{M}_{\odot}]< 10^{14} $, the samples suggest that the galaxies are predominantly red, although the distribution of the filament outskirts shows a higher dispersion with respect to the filament distribution.
Furthermore, these surrounding regions do not have halos with masses higher than $ 10^{14}\ h^{-1} \text{M}_{\odot}$, as mentioned earlier. 

\begin{figure}
    \centering
    {
        \includegraphics[width=0.47\textwidth]{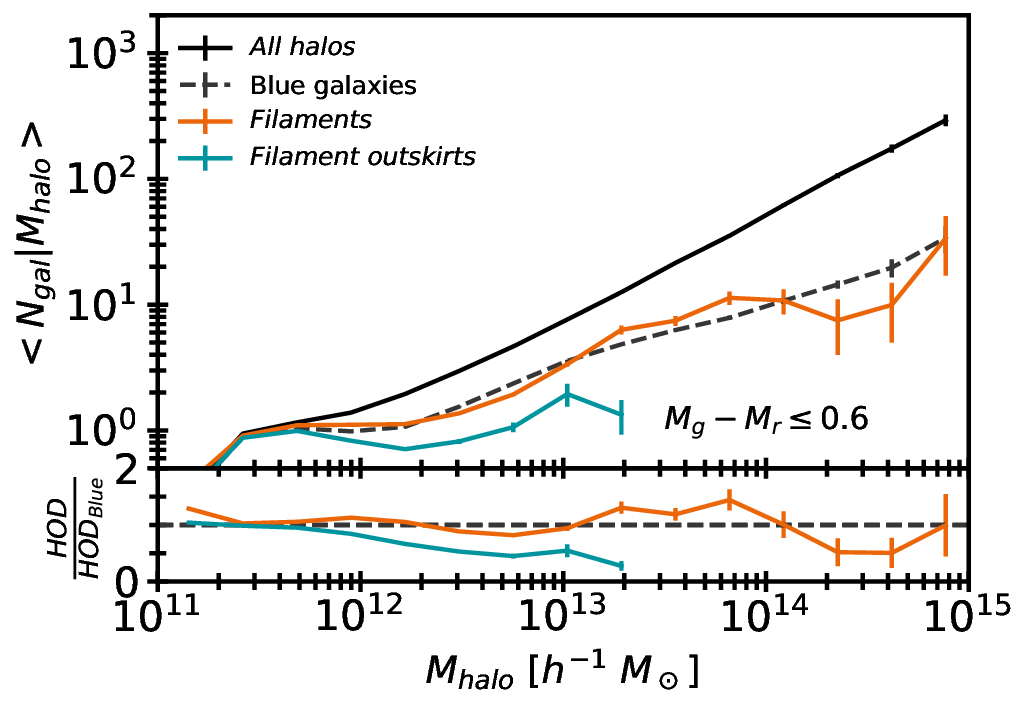}
    }
    {
        \includegraphics[width=0.47\textwidth]{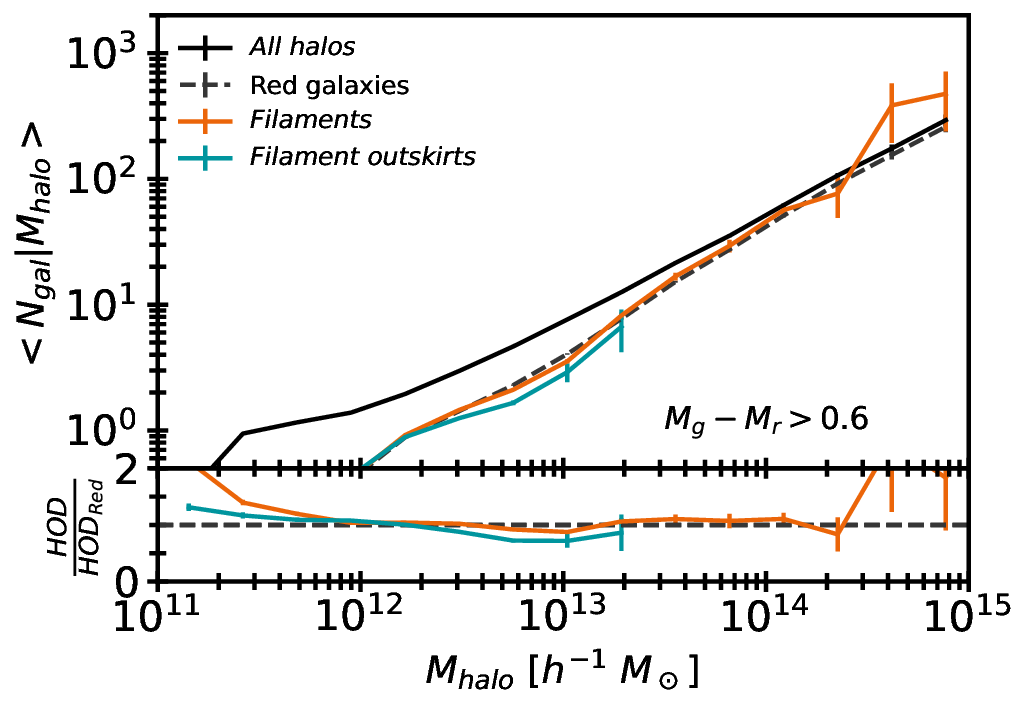}
    }
    \caption{Panel shows the measured HOD with respect to the galaxy colours for the total (black), filaments (orange) and filament outskirts (light-blue) samples for galaxies with $\text{M}_\text{r} - 5\text{log}_{10}(\text{h}) \leq -17$. 
    The dashed lines represent the HODs for the total of blue (upper) and red (lower) galaxies.  The lower sub-panels show the ratio of the filaments and filament outskirts samples to the total number of blue and red galaxies, respectively.}
    \label{fig:hod_fil_color}
\end{figure}

Progressing with the analysis of HODs in terms of galaxy properties, in Fig. \ref{fig:hod_fil_color}, we study the HODs for blue ($\text{M}_{\text{g}} - \text{M}_{\text{r}} \leq 0.6$) and red ($\text{M}_{\text{g}} - \text{M}_{\text{r}} > 0.6$) galaxies for each sample. 
As expected, the low-mass halos are preferentially populated by blue galaxies, in contrast to the high-mass halos dominated by red galaxies.
The tendencies of the red galaxies for the two samples are close to each other and lie on the distribution of the total red galaxies. 
This suggests that the halo occupation by red galaxies is similar in both environments.
On the other hand, an apparent disparity is observed when blue galaxies are considered. 
While the filaments sample tends to approximate the total blue sample within the error bars, the  surrounding regions, corresponding to the filaments outskirts, show a slight decrease in blue galaxies as we consider more massive halos. 

\begin{figure}
    \centering
    \includegraphics[width=0.5%
    \textwidth]
    {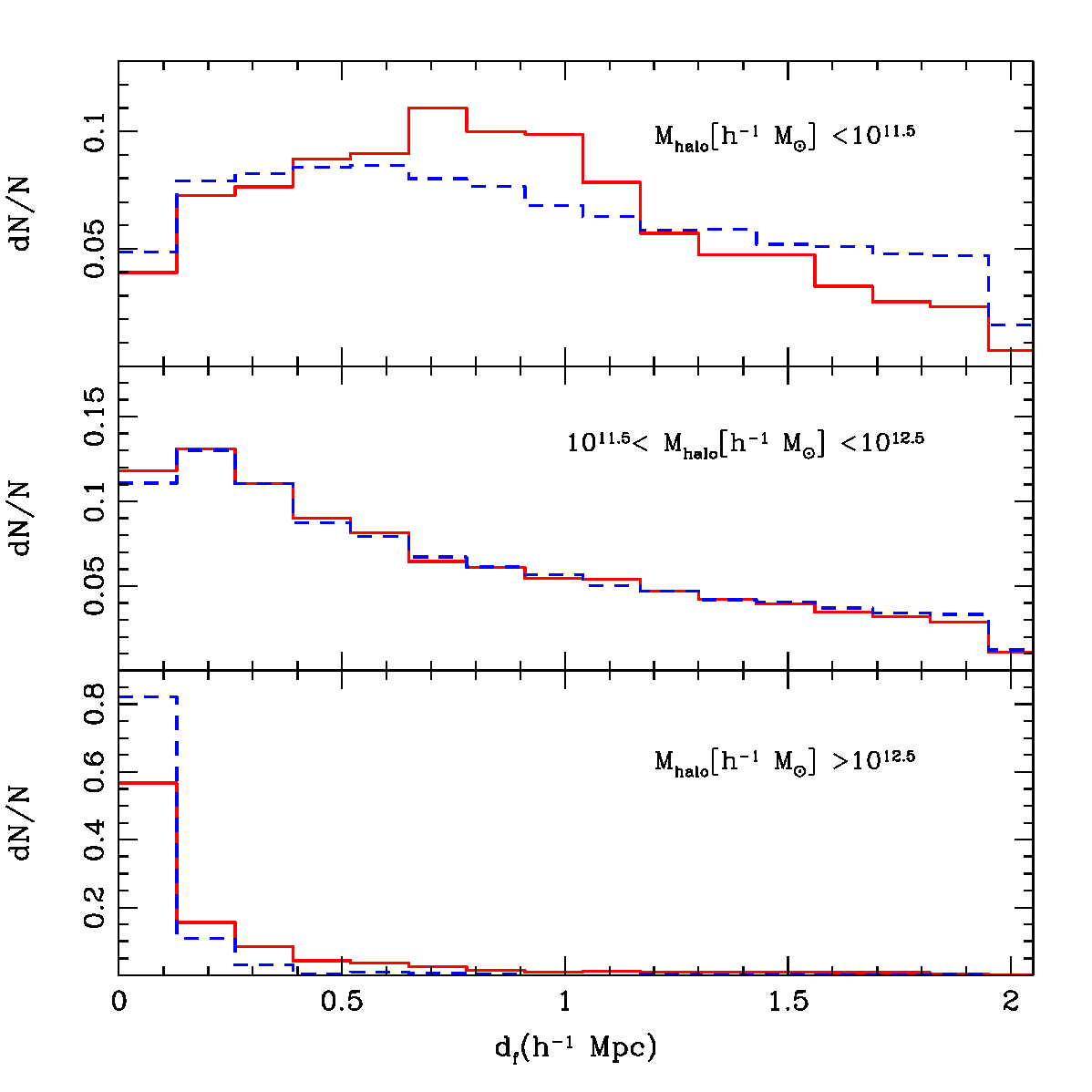}
    \caption{Normalised distributions of red (solid red lines) and blue (dashed blue lines) galaxies are presented for the low, intermediate and high halo mass range in the top, middle and bottom panels respectively.}
    \label{fig:hist_disteje_masah}
\end{figure}

The results of \cite{Kraljic2018} show that although the colour density segregation is milder in filaments than in regions of high peak density, it is still present, and it can be observed that red and passive galaxies are closer to the filaments than their star-forming and blue counterparts, regardless of the distance to the nodes.
To understand the lack of blue galaxies in our filament outskirts for the intermediate halo masses, we examined the colour-density segregation for the defined blue and red galaxy samples considering three halo mass ranges, selected to visualise the colour-density tendencies in different halo mass ranges.
In Fig. \ref{fig:hist_disteje_masah}, the normalised distributions of blue and red galaxies in relation to the distance from the filament axis can be observed for low ($\text{M}_{\text{halo}} \leq 10^{11.5}\, h^{-1}\, M_{\odot}$), intermediate ($\text{M}_{\text{halo}} \in (10^{11.5}, 10^{12.5})\, h^{-1}\,\text{M}_{\odot}$) and high ($\text{M}_{\text{halo}} > 10^{12.5}\, h^{-1}\text{M}_{\odot}$) mass range, respectively.
For the low-mass halo region, close to the filament axis, the number of galaxies (both blue and red) decreases, making the morphological segregation unclear, while this effect becomes more pronounced towards the outer regions of the filament. 
Here, the proportion of blue galaxies increases with distance from the filament axis, in contrast to the proportion of red galaxies.
For intermediate halo-masses, the colour-density segregation is still present but weaker. However, in the high halo-mass range, an excess of blue galaxies is observed close to the filament axis, and the trend is reversed.
Then, the blue galaxies diminishing in high-mass halos associated with the outskirts could be explained by most galaxies in these environments being reddened by local halo effects.

%


\begin{figure}
\centering
    {
        \includegraphics[width=0.47\textwidth]{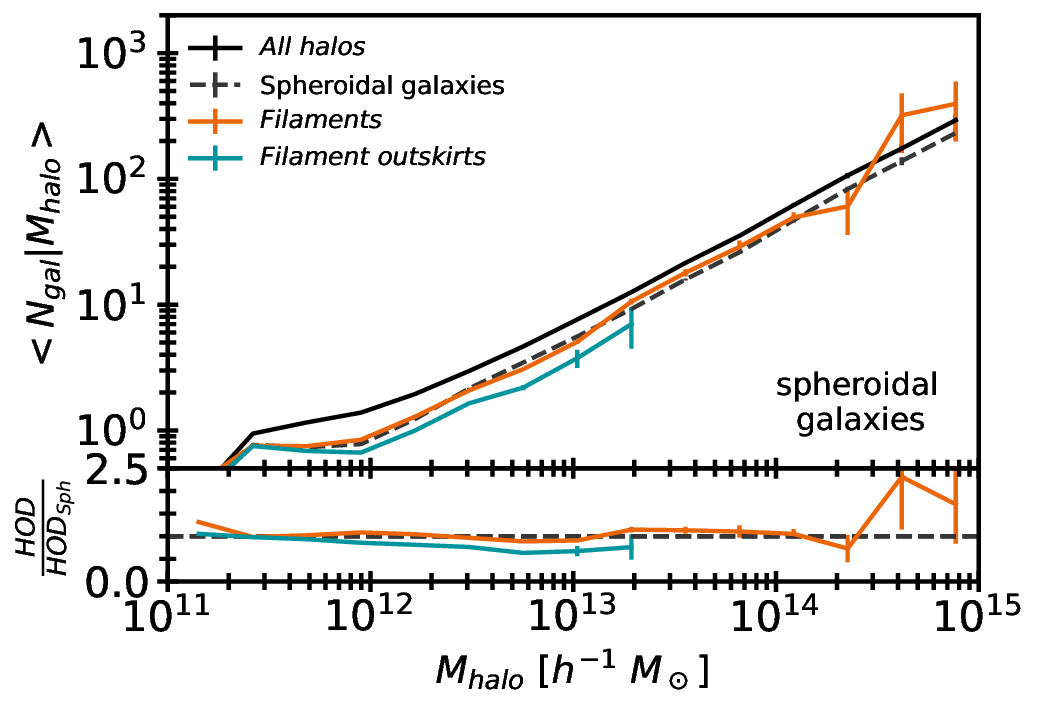}
    }
    {
        \includegraphics[width=0.47\textwidth]{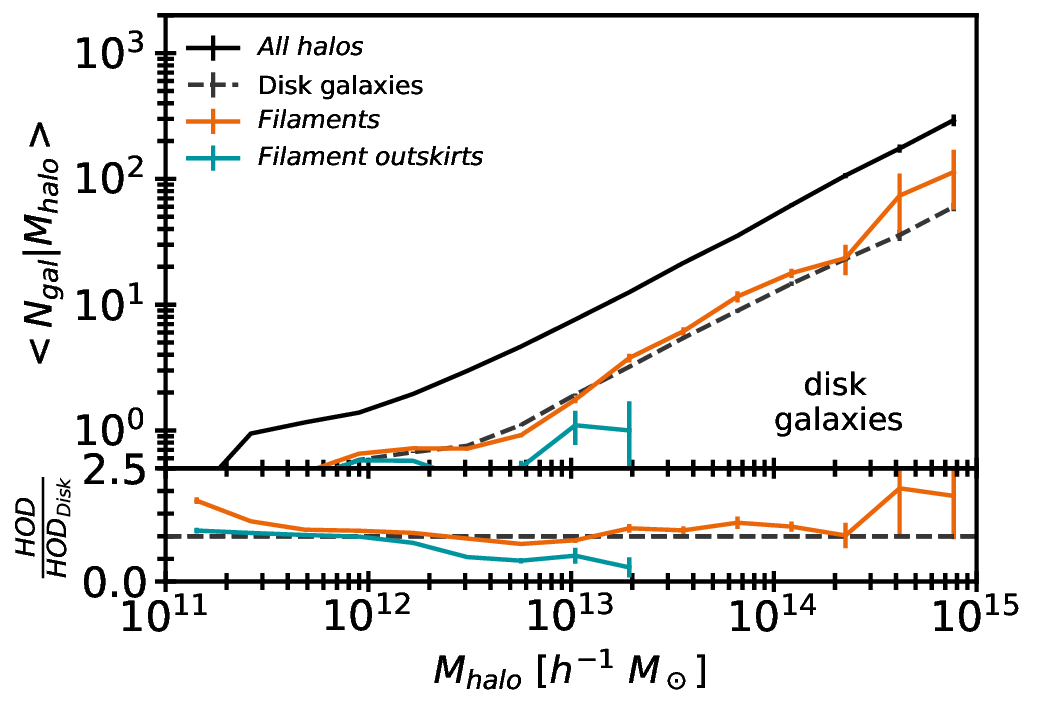}
    }     
    \caption{Panel shows the measured HOD with respect to the subhalo morphology for the total (black), filaments (orange) and filament outskirts (light-blue) samples for subhalos with $\text{M}_\text{r} - 5\text{log}_{10}(\text{h}) \leq -17$. The dashed lines represent the HODs for the total of spheroidal (upper) and disk (lower) subhalos. Lower sub-panels show the ratio between the filaments and filament outskirts samples with the total of spheroidal and disk subhalos, respectively.}
    \label{fig:hod_fil_morf}
\end{figure}

\begin{figure}
\centering
    {
        \includegraphics[width=0.47\textwidth]{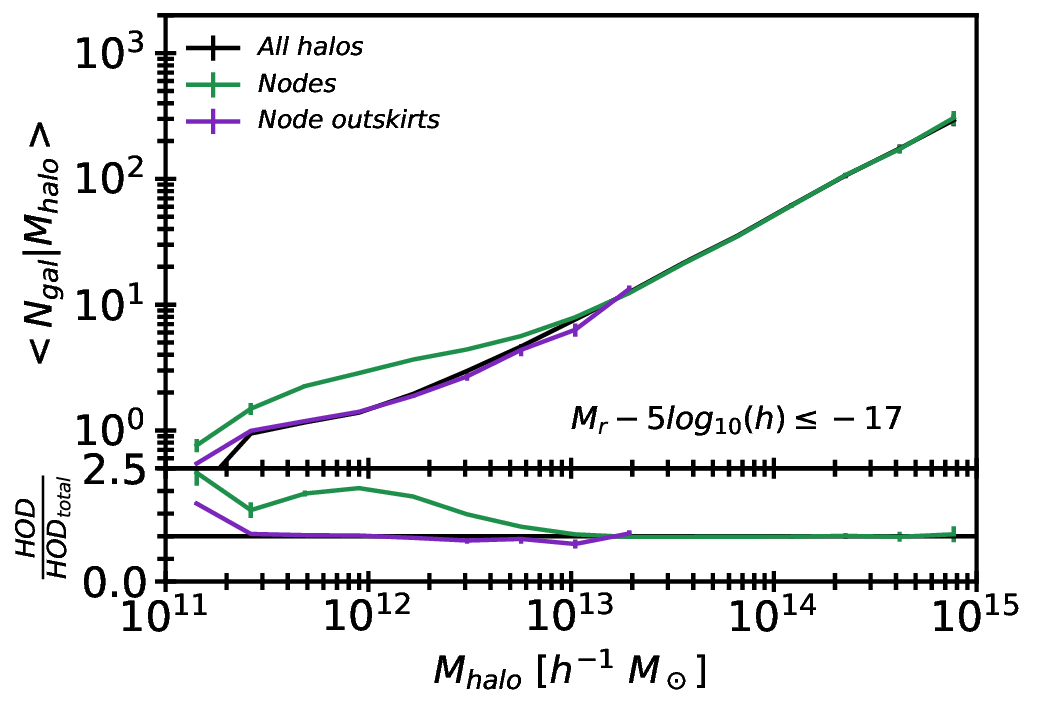}
    }
    {
        \includegraphics[width=0.47\textwidth]{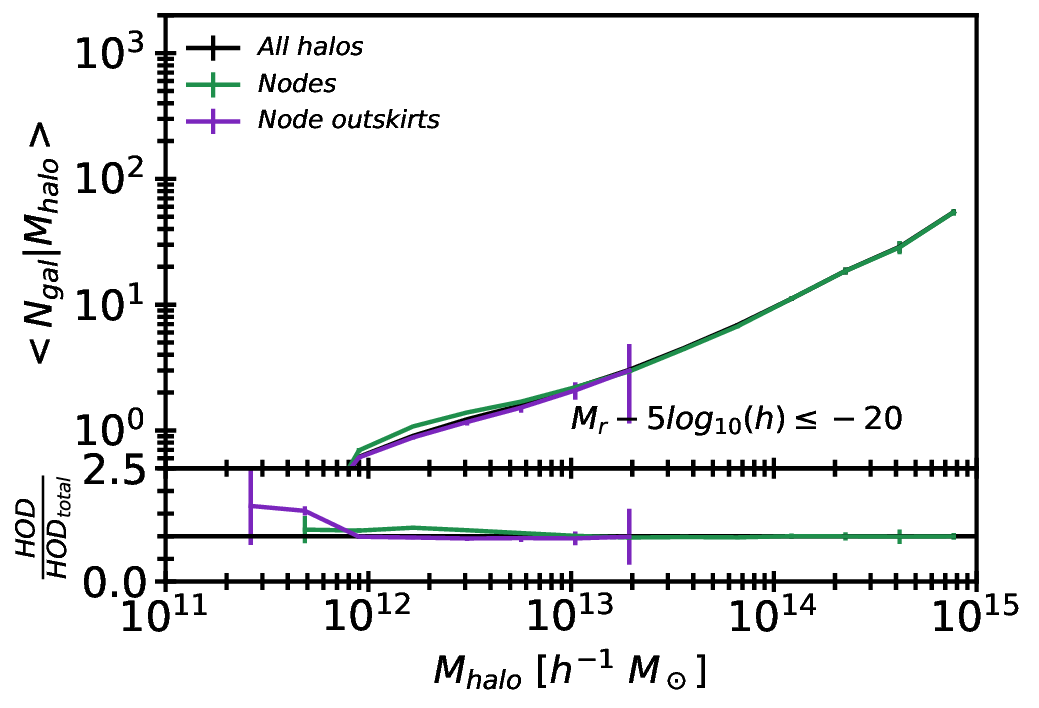}
    }
    \caption{HODs measured for the total (black), nodes (green) and node outskirts (purple) samples for subhalos with $\text{M}_\text{r} - 5\text{log}_{10}(\text{h}) \leq -17$ (top panel) and $\text{M}_\text{r} - 5\text{log}_{10}(\text{h}) \leq -20$ (bottom panel), respectively. Lower sub-panels show the ratio between the nodes and node outskirts samples with the overall HOD.}
    \label{fig:hod_nod}
\end{figure}

\begin{figure*}
\centering
    {
        \includegraphics[width=0.8\textwidth]{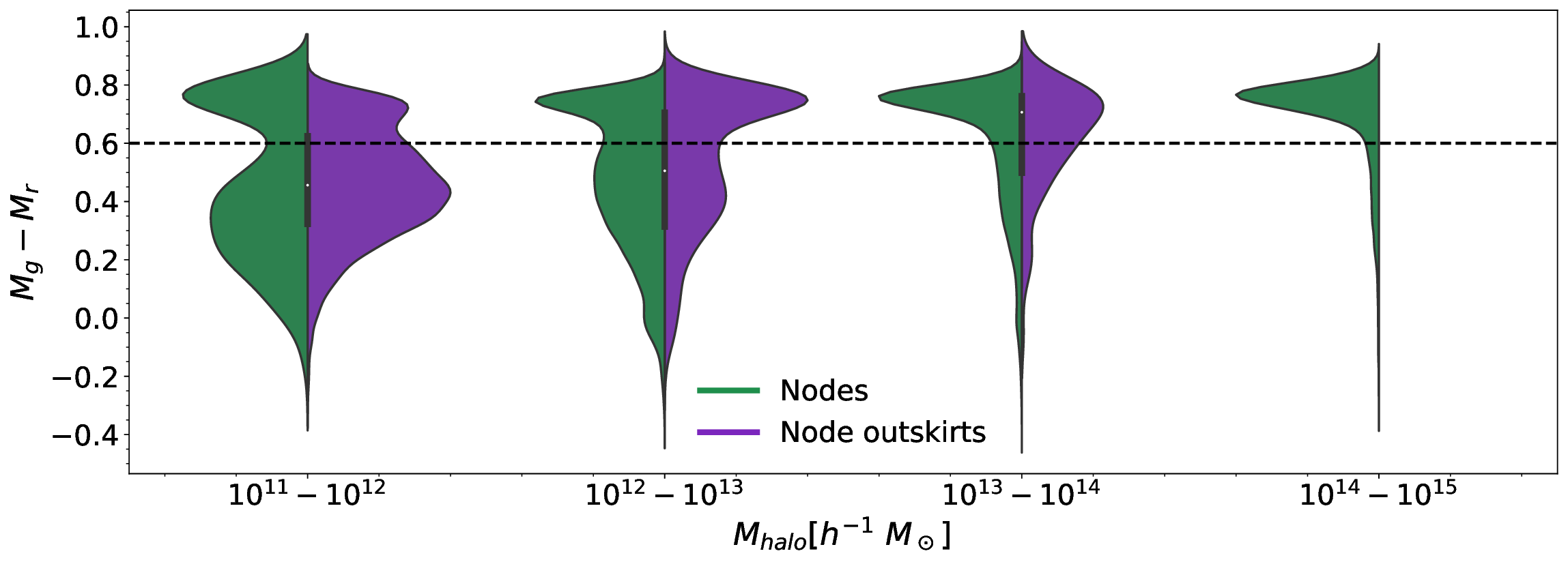}
    }
    \caption{Galaxy colour distributions in halo mass bins for nodes (green) and node outskirts (purple) samples.}
    \label{fig:distr_gr_masas_nod}
\end{figure*}

\begin{figure}
\centering
    {
        \includegraphics[width=0.47\textwidth]{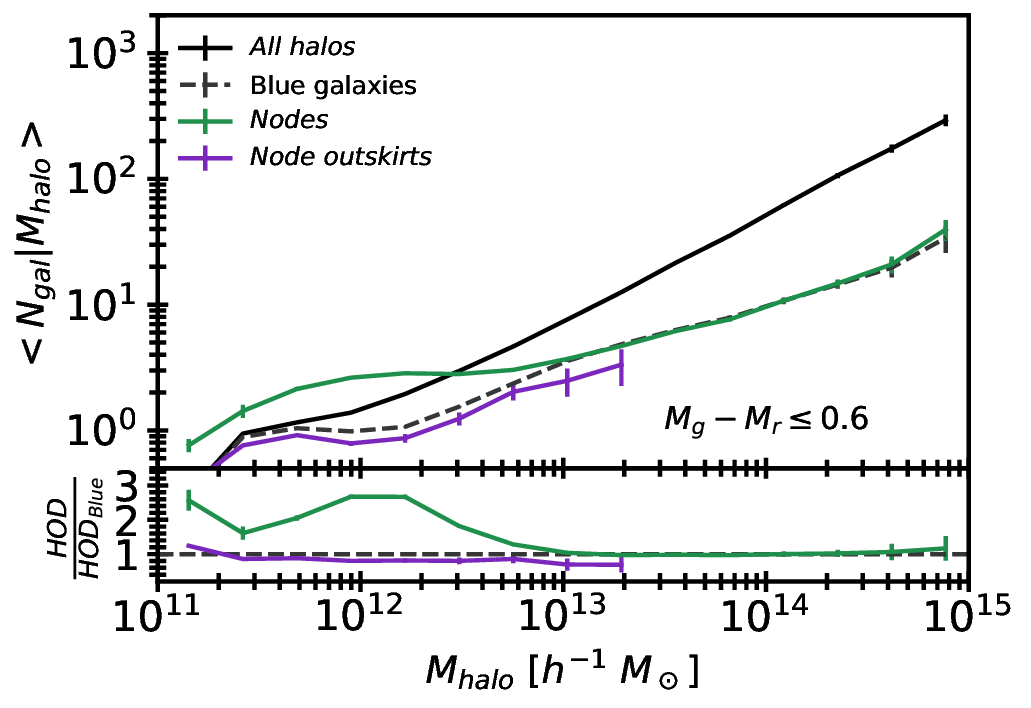}
    }
    {
        \includegraphics[width=0.47\textwidth]{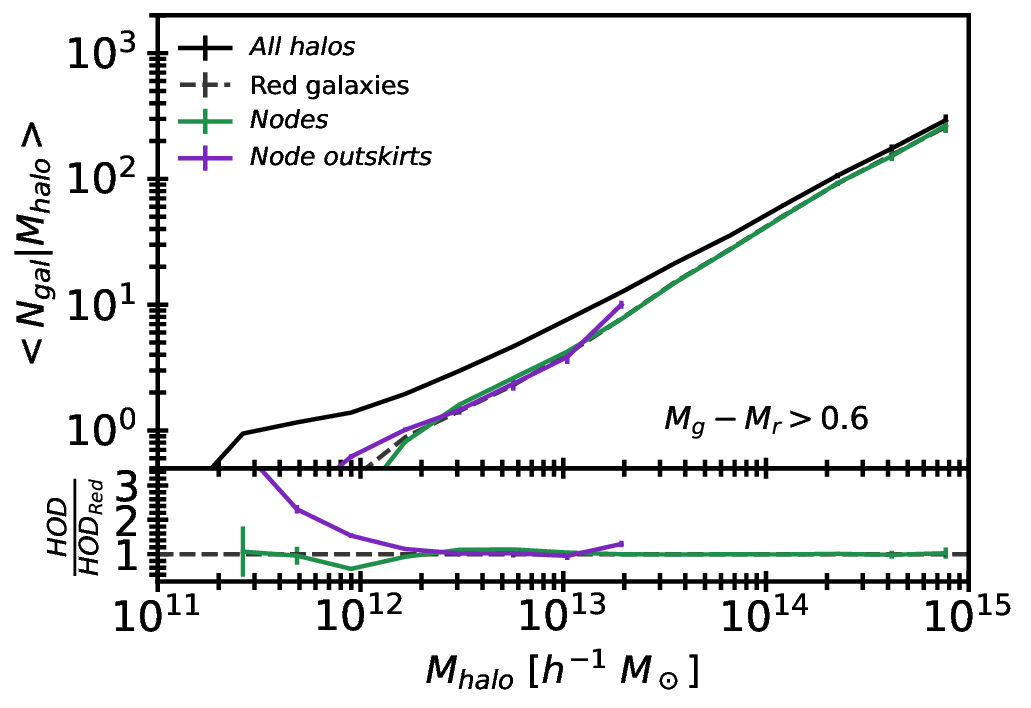}
    }
    \caption{
        The main panel shows the measured HODs with respect to the galaxy colours for three samples: total (black), nodes (green), and node outskirts (purple), considering the subhalos with a magnitude of $\text{M}_\text{r} - 5\text{log}_{10}(\text{h}) \leq -17$. 
        The dashed lines in the upper and bottom panels represent the HODs for the total blue and red galaxies, respectively. Bellow each panel, we plot the ratio of the nodes and node outskirts samples with respect to the total number of blue and red galaxies    
    }
    \label{fig:hod_nod_color}
\end{figure}

\begin{figure}
\centering
    {
        \includegraphics[width=0.47\textwidth]{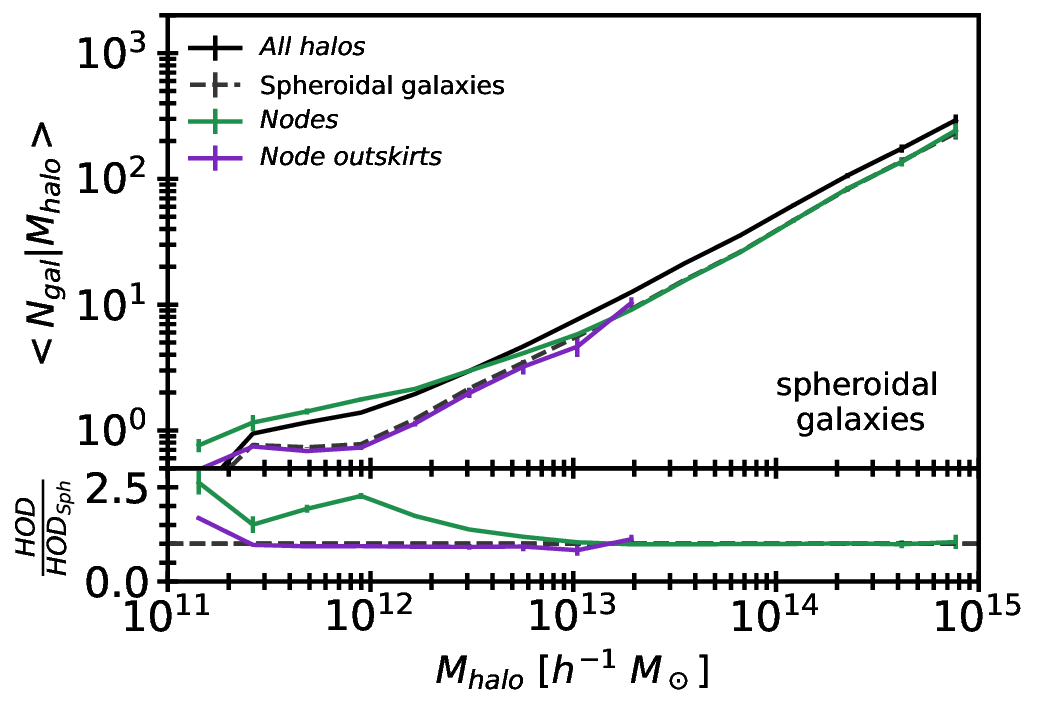}
    }
    {
        \includegraphics[width=0.47\textwidth]{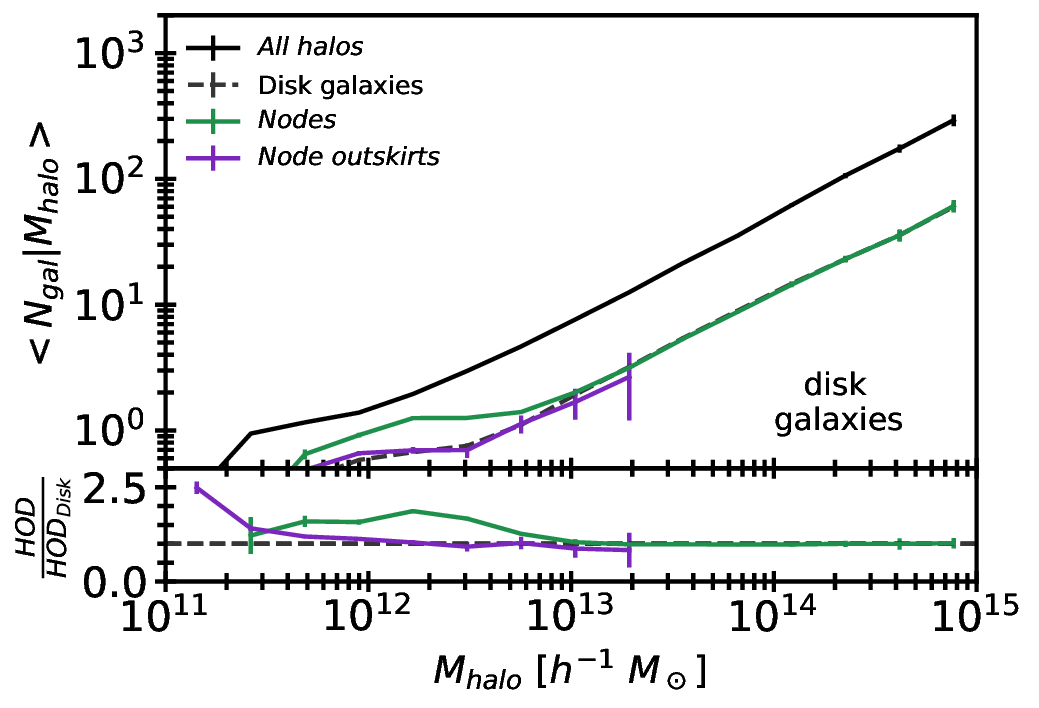}
    }      
    \caption{Panel shows the measured HODs with respect to the subhalo morphology for the total (black), nodes (green) and node outskirts (purple) samples for subhalos with $\text{M}_\text{r} - 5\text{log}_{10}(\text{h}) \leq -17$. The dashed lines represent the HODs for the total of spheroidal (upper) and disk (lower) subhalos. Lower sub-panels show the ratio between the nodes and node outskirts samples with the total of spheroidal and disk subhalos, respectively.}
    \label{fig:hod_nod_morf}
\end{figure}

To conclude the analysis of the HOD taking into account the galaxy properties we show the distributions with respect to morphology, as described in previous section, in Fig. \ref{fig:hod_fil_morf}.
%
The HODs for both filaments samples are comparable to the global trends, and although the HODs for filament outskirts are lower for both morphologies, the halo occupation is significantly lower for disk galaxies. 
These results are consistent with those obtained for galaxy colour.

The HODs measurements for the filaments sample, taking into account the studied galaxy properties, show that the halo occupancy in these environments is quite similar to the general trend, while halos belonging to the filament outskirts show HODs up to $0.5$ times lower than the general trend, despite the considered property.
It is important to note that our definition of filament environment does not take into account the filament radius according to its density profile. 
Moreover, the results of this subsection suggest that the colour-density segregation is fulfilled for low-mass halos (except for the regions close to the filamentary axis, where the number of galaxies is low and the result is therefore noisy), while decreasing for higher-mass halos, where the effect of local halo influences dominates over the proximity of the filamentary axis.


\subsection{Nodes}\label{ssec:Nodes}

For the sake of consistency, we apply the analysis of the previous subsection to 
the environments corresponding to the nodes sample. To account for  HOD variations in the environment associated with such cosmic web structures, we have also constructed a sample of the node periphery, called node outskirts, by selecting halos with $1 < \text{d}_{\text{n}}\ [\text{R}_{\text{200}} ] \leq 3$ to obtain halos in the node vicinity.
The number of halos and subhalos can be found in Table \ref{tab:num_halos}.

In Fig. \ref{fig:hod_nod}, we show the HODs for subhalos, with magnitude threshold $\text{M}_\text{r} - 5\text{log}_{10}(h) \leq -17 \ \text{and} -20$, for the total, nodes and node outskirts samples.
For the first magnitude threshold it is observed that the outskirts sample lies above the total sample, although in a smaller halo mass range due to the absence of massive halos in these regions. That is, the node outskirts does not seem to have any influence on the halo occupancy.
In addition, it can be observed that the behaviour within the range of masses covered is quite similar to that of the filament sample in the previous subsection.
On the other hand, the node sample presents a remarkable difference for halo masses $M_{\rm halo} < 10^{13} h^{-1} M_{\odot}$, where the HOD measured for the nodes is $2.5$ times higher than the trend of the other samples.
This result suggests an excess of faint galaxies in the low-mass halo population within nodes,
indicating that the low-mass halos are still in formation.
Later forming halos exhibit more recent accretion events, resulting in an excess of substructure. In contrast, early forming halos contain fewer subhalos due to more time for tidal disruption interactions or merging with the host halo via dynamical friction \citep{Tinker2021}.

Measuring the HOD using only the brightest galaxies, we find that the difference becomes negligible. Thus, the location of the host halo does not affect the occupation of the halo by a bright galaxy.
As the largest difference in the measured HODs was detected at the magnitude threshold $\text{M}_\text{r} - 5\text{log}_{10}(h) \leq -17$, we set this limit for the next analysis.

In line with the previous subsection, we examine the galaxy colour distributions for both environments in halo mass bins.
Fig. \ref{fig:distr_gr_masas_nod} shows that low-mass halos in nodes exhibit a distinct bimodal colour distribution, with the red component becoming more significant as mass increases, until the blue component disappears completely for massive halos.
This outcome is in line with galaxy colour change over time. In high-mass halos in nodes, galaxies are more evolved as they have had more time to deplete their gas reserves and cease star formation, becoming redder. However, for the lower halo mass in nodes, many galaxies are still being accreted, leading to a population of young objects observed in the bimodal distribution.
Nevertheless, the low-mass halos within the node outskirts show a distinct colour bimodality with an overpopulation of objects in the green valley. Also, the red component becomes more significant with increasing mass.
The last mass bin is empty, reflecting the reason for the mass limit in the HODs for these environments.
%
%
These differences between galaxy colours at the nodes and at their periphery could be explained by environmental morphological segregation \citep{Dominguez2001}. Then, the node outskirts allows to find galaxies relatively uninfluenced by the dominant effects within the nodes.
In addition, \cite{Martinez2008} found that colour is the property that depends most on the distance from the centre of the cluster.
In this sense, the overpopulation of galaxies in the green valley in halos belonging to node outskirts sample could be related to the fact that 
the colour of these objects is affected by their proximity to the node.

Following the same line as the one in the previous subsection in  Fig. \ref{fig:hod_nod_color}, we show the HODs considering the galaxy colours. 
In the case of blue galaxies (upper panel), the nodes sample shows an excess of blue galaxies for halos with masses lower than $10^{13} h^{-1} M_{\odot}$ while the halos inhabiting in the node outskirts exhibit a HOD slightly lower than the average HOD, within the error bars.
For the red galaxies (lower panel), the trends for the HODs for both samples (nodes and their outskirts) are similar and lie close to the distribution of the total red galaxies. 
This result suggests that the occupation of the halo by blue galaxies depends on their location respect to the node centre, while red galaxies populate the halos regardless of their environment.
%


Finally, in Fig. \ref{fig:hod_nod_morf} we show the HODs as a function of the galaxy morphology. 
The HODs calculated with spheroidal galaxies indicate that low-mass halos in nodes have an excess of early-type galaxies with respect to the total sample.
In the intermediate mass range, the distribution decreases until it overlaps with the HOD for a total of spheroidal galaxies.
On the other hand, the node outskirts sample coincides with the total early-type galaxies distribution, regardless of halo mass.
For disk galaxies, the HOD for node halos is higher than the total distribution of disk galaxies in the low-mass region, but both samples overlap in the high-mass halo region.
The HOD for halos in the vicinity of nodes is close to the total distribution of disk galaxies, with a slight difference in the low-mass halo range.
Galaxy accretion in these regions is likely to result in many interactions that affect galaxy morphology.
However, 
the morphology of galaxies in low-mass halos is not an entirely reliable parameter.

The HODs  measured for the nodes sample show an excess of faint and blue 
galaxies for low-mass halos. 
Regarding to the morphology, the HODs show an excess of spheroidal and disk galaxies in the lower mass range.
These trends could be related to the measurement of the circularity parameter for low-mass galaxies, which have few stars to obtain an accurate result \citep{Abadi2003}.
On the other hand, the halo occupancy in the node outskirts is quite similar to the overall trend, regardless of the property considered.



\section{Summary and Conclusions} \label{sec:conclusions}

We have performed the Halo Occupation Distribution (HOD) measurements for halos derived from Groupcat at $z = 0.00$ from \textsc{Illustris TNG300-1} simulations.
Firstly, we separated halos belonging to filaments and nodes, taking into account the distance to the node and to the filament axis.
The nodes sample was constructed by considering halos with distance to the node $\text{d}_{\text{n}} \leq 1\ \text{R}_{\text{200}}$ and for the filaments sample we selected halos with distance to the node $\text{d}_{\text{n}} > 1\ \text{R}_{\text{200}}$ and distance to the filament axis $\ \text{d}_{\text{f}} \leq 1\ \mathrm{h^{-1}} \mathrm{Mpc}$.

We compute the mean number of galaxies per halo mass bin for the total, nodes and filaments samples considering four magnitude thresholds: $\text{M}_\text{r} - 5\text{log}_{10}(h) \leq -17, -18, -19\ \text{and} -20$,  following the line of \cite{Alfaro2020}. We find that the HOD decreases when we consider increasingly brighter galaxies.
Furthermore, the HOD for halos in filaments is similar to that shown for the total sample, regardless of galaxy luminosity.
This may suggest that the filamentary environment does not play a significant role in the occupancy of galaxies in halos, since the halo mass of the host galaxy could be the primary factor influencing the properties of the galaxy, as suggested by \cite{WhiteRees78}.
On the other hand, the nodes sample shows a significant excess of faint galaxies at lower halo masses, which decreases for bright galaxies.
Low-mass halos could have low rates of interaction and merging or may have undergone recent accretion processes. Therefore, these areas would favour the presence of faint galaxies.
Since the differences in the HOD measurements are more pronounced for the $\text{M}_\text{r} - 5\text{log}_{10}(h) \leq -17$ magnitude limit, we have chosen this cut-off magnitude for a more detailed analysis in the following. 

The distribution of colours in the total sample indicates that galaxies in low-mass halos are bluer. As mass increases, the galaxies become redder, which is consistent with the findings of \cite{Nelson2018}. The filaments and nodes samples follow the general trend, except for low-mass halos associated with nodes, which have a bimodal colour distribution. These low-mass nodes may represent local density maxima in the cosmic web that contain groups of galaxies. The bimodal distribution may indicate physical processes that cause galaxies to redden before they reach the virial radius \citep{Blanton2003,Kraljic2018}. Later, we measure the HODs taking into account several galaxy properties such as colour and morphology in filaments, nodes and their surrounding regions.

Then, in order to study the variations associated with the distance to filamentary structures we define a filament outskirts sample by selecting halos with $\text{d}_{\text{n}} > 1 \text{R}_{\text{200}}$ and $1  < \text{d}_{\text{f}}\ [h^{-1} \text{Mpc}] \leq 2 $. 
The HODs for subhalos with $\text{M}_\text{r} - 5\text{log}_{10}(h) \leq -17$ show that the filaments follow the overall trend, while the filament outskirts sample with $\text{M}_{\text{halo}} > 10^{12}\ h^{-1} \text{M}_{\odot}$ is 0.5 times lower than the total, suggesting that these less dense environments are less likely to be inhabited by galaxies.
The HODs measured using the brighter galaxies show similar results, with both samples lying on the total HOD, despite the error bars.
This result is in agreement with \cite{Alfaro2020}, who found similar trends for even less dense regions such as cosmic voids.
Looking at galaxy colours, despite the absence of the more massive halos in the outskirts, the galaxy colour distributions are quite similar in both samples although,  in the range of $10^{13}  < \text{M}_{\text{halo}} [h^{-1} \text{M}_{\odot}] < 10^{14}$ the distribution of the filament outskirts shows a higher dispersion with respect to the filament distribution.
The calculated HODs, discriminating the galaxy colours, show that the halo occupation of red galaxies is quite similar in both environments, while blue galaxies are less likely to be found in the outskirts than in the filaments for halo masses $\text{M}_{\text{halo}} > 10^{12}\ h^{-1} \text{M}_{\odot}$. In addition, the morphology analysis reveals that the halos in the filaments are populated by disk and spheroidal galaxies, which is consistent with the overall trend. However, the outskirts of the filaments are generally devoid of galaxies, although they contain significantly fewer disk galaxies.
To comprehend this outcome we analyse the normalised distributions of blue and red galaxies with respect to the  distance from the filament axis for different halo masses. 
We observed a pronounced colour-density segregation, except near the filament axis, for $\text{M}_{\text{halo}} < 10^{11.5}\ h^{-1} \text{M}_{\odot}$, in agreement with \citet{Kraljic2018}. 
As the halo mass increases, the degree of segregation weakens and for $\text{M}_{\text{halo}} > 10^{12.5}\ h^{-1} \text{M}_{\odot}$ 
the trend is reversed.
The absence of blue galaxies in high-mass halos situated in the filament outskirts may be attributed to a halo effect, such as the halo formation time \citep{Tinker2021}, or because the local dense environment has a greater influence on the galaxy properties \citep{WhiteRees78} than the global environment.

%

Following the study in the filamentary regions, we examined the nodes and their surrounding regions.
The node outskirts sample was built considering halos with $1 < \text{d}_{\text{n}}\ [\text{R}_{\text{200}} ]  \leq 3$.
Taking into account the magnitude limit $\text{M}_\text{r} - 5\text{log}_{10}(h) \leq -17$, we observe a remarkable excess of faint galaxies in the halo occupation for halos with masses lower than $ 10^{13} h^{-1} \text{M}_{\odot}$, while the outskirts follow the global trend.
For galaxies with $\text{M}_\text{r} - 5\text{log}_{10}(h) \leq -20$, the HODs for both samples are similar to the total sample.
The galaxy colour distributions for low-mass halos are bimodal, and with increasing mass the red component becomes more significant in both samples, as stated by the environmental morphological segregation \citep{Dominguez2001}, although there are no massive halos in the outskirts.
The HODs with respect to galaxy colours show an excess of blue galaxies for halos with masses lower than $ 10^{13}\ h^{-1} \text{M}_{\odot}$ in the nodes, while the node outskirts show a halo occupation slightly lower than the total sample.
Regarding morphology, the HODs for nodes show an excess of both, disk and bulge galaxies, for nodes with masses lower than $ 10^{13}\ h^{-1} \text{M}_{\odot}$, while the halos within the node outskirts have an occupancy comparable to the global trend.
%

The halo occupation in filaments appears to be independent of the properties of the galaxies inhabiting these environments and similar to
the global tendency. Instead of that the outskirts of the filaments show a lower halo occupancy.
Regarding the nodes, the HODs for low-mass halos show a significant difference with respect to the other samples. In this sense, the nodes could represent structures where the galaxies fall through the filaments, and they could be at different stages of evolution depending on their mass. 
These results could be related to the halo assembly bias, as proposed by \cite{Ramakrishnan2019} and \cite{Mansfield2020}, who consider this important issue for the formation and evolution of the large structures of the Universe.
Also, works such as \cite{Borzyszkowski2017} point out that such a bias is a product of the action of large structures in the Universe that cause the extinction of halo growth through tidal forces.
In particular, they claim that ``stalled'' halos would be found at the nodes. The changes we observe in the HOD could result from just such a difference in the formation history of halos.

The analysis carried out in this study provides insights into how galaxies populate the halos, and how this depends on the mass of the halo and its position within the cosmic web.
The study of the evolution of HODs leads to a very interesting topic of analysis, as suggested by \cite{Contreras2023}. In a forthcoming work, we will then assess the evolution of HODs in the large-scale environments given by nodes and filamentary structures.


\section*{Acknowledgements}
We thank the referee, for providing us with helpful comments and suggestions that improved this paper.
This work was supported in part by the Consejo Nacional de Investigaciones Cient\'ificas y T\'ecnicas de la Rep\'ublica Argentina (CONICET) and the Consejo Nacional de Investigaciones Cient\'ificas, T\'ecnicas y de Creaci\'on Art\'istica de la Universidad Nacional de San Juan (CICITCA).
The authors would like to thank the \textsc{IllustrisTNG} team for making their data available to the public.
FR would like to acknowledge support from the ICTP through the Junior Associates Programme 2023-2028.
\section*{Data Availability}

The simulation data underlying this article are publicly available at
the TNG website. The data results arising from this work will be
shared on reasonable request to the corresponding authors.



\bibliographystyle{mnras}
\bibliography{bibliography} 

\begin{thebibliography}{}
\makeatletter
\relax
\def\mn@urlcharsother{\let\do\@makeother \do\$\do\&\do\#\do\^\do\_\do\%\do\~}
\def\mn@doi{\begingroup\mn@urlcharsother \@ifnextchar [ {\mn@doi@}
  {\mn@doi@[]}}
\def\mn@doi@[#1]#2{\def\@tempa{#1}\ifx\@tempa\@empty \href
  {http://dx.doi.org/#2} {doi:#2}\else \href {http://dx.doi.org/#2} {#1}\fi
  \endgroup}
\def\mn@eprint#1#2{\mn@eprint@#1:#2::\@nil}
\def\mn@eprint@arXiv#1{\href {http://arxiv.org/abs/#1} {{\tt arXiv:#1}}}
\def\mn@eprint@dblp#1{\href {http://dblp.uni-trier.de/rec/bibtex/#1.xml}
  {dblp:#1}}
\def\mn@eprint@#1:#2:#3:#4\@nil{\def\@tempa {#1}\def\@tempb {#2}\def\@tempc
  {#3}\ifx \@tempc \@empty \let \@tempc \@tempb \let \@tempb \@tempa \fi \ifx
  \@tempb \@empty \def\@tempb {arXiv}\fi \@ifundefined
  {mn@eprint@\@tempb}{\@tempb:\@tempc}{\expandafter \expandafter \csname
  mn@eprint@\@tempb\endcsname \expandafter{\@tempc}}}

\bibitem[\protect\citeauthoryear{{Abadi}, {Navarro}, {Steinmetz}  \&
  {Eke}}{{Abadi} et~al.}{2003}]{Abadi2003}
{Abadi} M.~G.,  {Navarro} J.~F.,  {Steinmetz} M.,   {Eke} V.~R.,  2003, \mn@doi
  [\apj] {10.1086/378316}, \href
  {https://ui.adsabs.harvard.edu/abs/2003ApJ...597...21A} {597, 21}

\bibitem[\protect\citeauthoryear{{Alfaro}, {Rodriguez}, {Ruiz}  \&
  {Lambas}}{{Alfaro} et~al.}{2020}]{Alfaro2020}
{Alfaro} I.~G.,  {Rodriguez} F.,  {Ruiz} A.~N.,   {Lambas} D.~G.,  2020,
  \mn@doi [\aap] {10.1051/0004-6361/201937431}, \href
  {https://ui.adsabs.harvard.edu/abs/2020A&A...638A..60A} {638, A60}

\bibitem[\protect\citeauthoryear{Alfaro, Ruiz, Luparello, Rodriguez  \&
  Lambas}{Alfaro et~al.}{2021}]{Alfaro2021}
Alfaro I.~G.,  Ruiz A.~N.,  Luparello H.~E.,  Rodriguez F.,   Lambas D.~G.,
  2021, Astronomy \& Astrophysics, 654, A62

\bibitem[\protect\citeauthoryear{{Alfaro}, {Rodriguez}, {Ruiz}, {Luparello}  \&
  {Lambas}}{{Alfaro} et~al.}{2022}]{Alfaro2022}
{Alfaro} I.~G.,  {Rodriguez} F.,  {Ruiz} A.~N.,  {Luparello} H.~E.,   {Lambas}
  D.~G.,  2022, \mn@doi [\aap] {10.1051/0004-6361/202243542}, \href
  {https://ui.adsabs.harvard.edu/abs/2022A&A...665A..44A} {665, A44}

\bibitem[\protect\citeauthoryear{{Arag{\'o}n-Calvo}, {van de Weygaert}, {Jones}
   \& {van der Hulst}}{{Arag{\'o}n-Calvo} et~al.}{2007}]{Aragon-Calvo_2007}
{Arag{\'o}n-Calvo} M.~A.,  {van de Weygaert} R.,  {Jones} B. J.~T.,   {van der
  Hulst} J.~M.,  2007, \mn@doi [\apjl] {10.1086/511633}, \href
  {https://ui.adsabs.harvard.edu/abs/2007ApJ...655L...5A} {655, L5}

\bibitem[\protect\citeauthoryear{{Artale}, {Zehavi}, {Contreras}  \&
  {Norberg}}{{Artale} et~al.}{2018}]{Artale2018}
{Artale} M.~C.,  {Zehavi} I.,  {Contreras} S.,   {Norberg} P.,  2018, \mn@doi
  [\mnras] {10.1093/mnras/sty2110}, \href
  {https://ui.adsabs.harvard.edu/abs/2018MNRAS.480.3978A} {480, 3978}

\bibitem[\protect\citeauthoryear{{Berlind} \& {Weinberg}}{{Berlind} \&
  {Weinberg}}{2002}]{Berlind2002}
{Berlind} A.~A.,  {Weinberg} D.~H.,  2002, \mn@doi [\apj] {10.1086/341469},
  \href {https://ui.adsabs.harvard.edu/abs/2002ApJ...575..587B} {575, 587}

\bibitem[\protect\citeauthoryear{{Berlind} et~al.,}{{Berlind}
  et~al.}{2003}]{Berlind2003}
{Berlind} A.~A.,  et~al., 2003, \mn@doi [\apj] {10.1086/376517}, \href
  {https://ui.adsabs.harvard.edu/abs/2003ApJ...593....1B} {593, 1}

\bibitem[\protect\citeauthoryear{{Blanton} et~al.,}{{Blanton}
  et~al.}{2003}]{Blanton2003}
{Blanton} M.~R.,  et~al., 2003, \mn@doi [\apj] {10.1086/375528}, \href
  {https://ui.adsabs.harvard.edu/abs/2003ApJ...594..186B} {594, 186}

\bibitem[\protect\citeauthoryear{{Blanton}, {Eisenstein}, {Hogg}, {Schlegel}
  \& {Brinkmann}}{{Blanton} et~al.}{2005}]{Blanton2005}
{Blanton} M.~R.,  {Eisenstein} D.,  {Hogg} D.~W.,  {Schlegel} D.~J.,
  {Brinkmann} J.,  2005, \mn@doi [\apj] {10.1086/422897}, \href
  {https://ui.adsabs.harvard.edu/abs/2005ApJ...629..143B} {629, 143}

\bibitem[\protect\citeauthoryear{{Bond}, {Kofman}  \& {Pogosyan}}{{Bond}
  et~al.}{1996}]{bond96}
{Bond} J.~R.,  {Kofman} L.,   {Pogosyan} D.,  1996, \mn@doi [\nat]
  {10.1038/380603a0}, \href
  {https://ui.adsabs.harvard.edu/abs/1996Natur.380..603B} {380, 603}

\bibitem[\protect\citeauthoryear{Borzyszkowski, Porciani, Romano-Diaz  \&
  Garaldi}{Borzyszkowski et~al.}{2017}]{Borzyszkowski2017}
Borzyszkowski M.,  Porciani C.,  Romano-Diaz E.,   Garaldi E.,  2017, Monthly
  Notices of the Royal Astronomical Society, 469, 594

\bibitem[\protect\citeauthoryear{{Cautun}, {van de Weygaert}, {Jones}  \&
  {Frenk}}{{Cautun} et~al.}{2014}]{Cautun2014}
{Cautun} M.,  {van de Weygaert} R.,  {Jones} B. J.~T.,   {Frenk} C.~S.,  2014,
  \mn@doi [\mnras] {10.1093/mnras/stu768}, \href
  {https://ui.adsabs.harvard.edu/abs/2014MNRAS.441.2923C} {441, 2923}

\bibitem[\protect\citeauthoryear{{Contreras} \& {Zehavi}}{{Contreras} \&
  {Zehavi}}{2023}]{Contreras2023}
{Contreras} S.,  {Zehavi} I.,  2023, \mn@doi [arXiv e-prints]
  {10.48550/arXiv.2305.19628}, \href
  {https://ui.adsabs.harvard.edu/abs/2023arXiv230519628C} {p. arXiv:2305.19628}

\bibitem[\protect\citeauthoryear{{Contreras}, {Baugh}, {Norberg}  \&
  {Padilla}}{{Contreras} et~al.}{2013}]{Contreras2013}
{Contreras} S.,  {Baugh} C.~M.,  {Norberg} P.,   {Padilla} N.,  2013, \mn@doi
  [\mnras] {10.1093/mnras/stt629}, \href
  {https://ui.adsabs.harvard.edu/abs/2013MNRAS.432.2717C} {432, 2717}

\bibitem[\protect\citeauthoryear{{Darvish}, {Mobasher}, {Sobral}, {Rettura},
  {Scoville}, {Faisst}  \& {Capak}}{{Darvish} et~al.}{2016}]{Darvish2016}
{Darvish} B.,  {Mobasher} B.,  {Sobral} D.,  {Rettura} A.,  {Scoville} N.,
  {Faisst} A.,   {Capak} P.,  2016, \mn@doi [\apj]
  {10.3847/0004-637X/825/2/113}, \href
  {https://ui.adsabs.harvard.edu/abs/2016ApJ...825..113D} {825, 113}

\bibitem[\protect\citeauthoryear{{Dom{\'\i}nguez}, {Muriel}  \&
  {Lambas}}{{Dom{\'\i}nguez} et~al.}{2001}]{Dominguez2001}
{Dom{\'\i}nguez} M.,  {Muriel} H.,   {Lambas} D.~G.,  2001, \mn@doi [\aj]
  {10.1086/319405}, \href
  {https://ui.adsabs.harvard.edu/abs/2001AJ....121.1266D} {121, 1266}

\bibitem[\protect\citeauthoryear{{Duckworth}, {Tojeiro}  \&
  {Kraljic}}{{Duckworth} et~al.}{2020a}]{Duckworth2020a}
{Duckworth} C.,  {Tojeiro} R.,   {Kraljic} K.,  2020a, \mn@doi [\mnras]
  {10.1093/mnras/stz3575}, \href
  {https://ui.adsabs.harvard.edu/abs/2020MNRAS.492.1869D} {492, 1869}

\bibitem[\protect\citeauthoryear{{Duckworth}, {Starkenburg}, {Genel}, {Davis},
  {Habouzit}, {Kraljic}  \& {Tojeiro}}{{Duckworth}
  et~al.}{2020b}]{Duckworth2020b}
{Duckworth} C.,  {Starkenburg} T.~K.,  {Genel} S.,  {Davis} T.~A.,  {Habouzit}
  M.,  {Kraljic} K.,   {Tojeiro} R.,  2020b, \mn@doi [\mnras]
  {10.1093/mnras/staa1494}, \href
  {https://ui.adsabs.harvard.edu/abs/2020MNRAS.495.4542D} {495, 4542}

\bibitem[\protect\citeauthoryear{{Einasto} et~al.,}{{Einasto}
  et~al.}{2008}]{Einasto2008}
{Einasto} M.,  et~al., 2008, \mn@doi [\apj] {10.1086/590374}, \href
  {https://ui.adsabs.harvard.edu/abs/2008ApJ...685...83E} {685, 83}

\bibitem[\protect\citeauthoryear{{Faltenbacher}, {Gottl{\"o}ber}, {Kerscher}
  \& {M{\"u}ller}}{{Faltenbacher} et~al.}{2002}]{Faltenbacher2002}
{Faltenbacher} A.,  {Gottl{\"o}ber} S.,  {Kerscher} M.,   {M{\"u}ller} V.,
  2002, \mn@doi [\aap] {10.1051/0004-6361:20021263}, \href
  {https://ui.adsabs.harvard.edu/abs/2002A&A...395....1F} {395, 1}

\bibitem[\protect\citeauthoryear{{Forero-Romero}, {Contreras}  \&
  {Padilla}}{{Forero-Romero} et~al.}{2014}]{Forero-Romero2014}
{Forero-Romero} J.~E.,  {Contreras} S.,   {Padilla} N.,  2014, \mn@doi [\mnras]
  {10.1093/mnras/stu1150}, \href
  {https://ui.adsabs.harvard.edu/abs/2014MNRAS.443.1090F} {443, 1090}

\bibitem[\protect\citeauthoryear{{Gal{\'a}rraga-Espinosa}, {Langer}  \&
  {Aghanim}}{{Gal{\'a}rraga-Espinosa} et~al.}{2022}]{GalarragaEspinosa2022}
{Gal{\'a}rraga-Espinosa} D.,  {Langer} M.,   {Aghanim} N.,  2022, \mn@doi
  [\aap] {10.1051/0004-6361/202141974}, \href
  {https://ui.adsabs.harvard.edu/abs/2022A&A...661A.115G} {661, A115}

\bibitem[\protect\citeauthoryear{{Gal{\'a}rraga-Espinosa}, {Garaldi}  \&
  {Kauffmann}}{{Gal{\'a}rraga-Espinosa} et~al.}{2023}]{GalarragaEspinosa2023}
{Gal{\'a}rraga-Espinosa} D.,  {Garaldi} E.,   {Kauffmann} G.,  2023, \mn@doi
  [\aap] {10.1051/0004-6361/202244935}, \href
  {https://ui.adsabs.harvard.edu/abs/2023A&A...671A.160G} {671, A160}

\bibitem[\protect\citeauthoryear{{Ganeshaiah Veena}, {Cautun}, {van de
  Weygaert}, {Tempel}, {Jones}, {Rieder}  \& {Frenk}}{{Ganeshaiah Veena}
  et~al.}{2018}]{Ganeshaiah_Veena2018}
{Ganeshaiah Veena} P.,  {Cautun} M.,  {van de Weygaert} R.,  {Tempel} E.,
  {Jones} B. J.~T.,  {Rieder} S.,   {Frenk} C.~S.,  2018, \mn@doi [\mnras]
  {10.1093/mnras/sty2270}, \href
  {https://ui.adsabs.harvard.edu/abs/2018MNRAS.481..414G} {481, 414}

\bibitem[\protect\citeauthoryear{{Ganeshaiah Veena}, {Cautun}, {Tempel}, {van
  de Weygaert}  \& {Frenk}}{{Ganeshaiah Veena}
  et~al.}{2019}]{GaneshaiahVeena2019}
{Ganeshaiah Veena} P.,  {Cautun} M.,  {Tempel} E.,  {van de Weygaert} R.,
  {Frenk} C.~S.,  2019, \mn@doi [\mnras] {10.1093/mnras/stz1343}, \href
  {https://ui.adsabs.harvard.edu/abs/2019MNRAS.487.1607G} {487, 1607}

\bibitem[\protect\citeauthoryear{{Gao} \& {White}}{{Gao} \&
  {White}}{2007}]{Gao2007}
{Gao} L.,  {White} S. D.~M.,  2007, \mn@doi [\mnras]
  {10.1111/j.1745-3933.2007.00292.x}, \href
  {https://ui.adsabs.harvard.edu/abs/2007MNRAS.377L...5G} {377, L5}

\bibitem[\protect\citeauthoryear{{Gao}, {Springel}  \& {White}}{{Gao}
  et~al.}{2005}]{Gao2005}
{Gao} L.,  {Springel} V.,   {White} S. D.~M.,  2005, \mn@doi [\mnras]
  {10.1111/j.1745-3933.2005.00084.x}, \href
  {https://ui.adsabs.harvard.edu/abs/2005MNRAS.363L..66G} {363, L66}

\bibitem[\protect\citeauthoryear{{Genel} et~al.,}{{Genel}
  et~al.}{2014}]{Genel2014}
{Genel} S.,  et~al., 2014, \mn@doi [\mnras] {10.1093/mnras/stu1654}, \href
  {https://ui.adsabs.harvard.edu/abs/2014MNRAS.445..175G} {445, 175}

\bibitem[\protect\citeauthoryear{{Genel}, {Fall}, {Hernquist}, {Vogelsberger},
  {Snyder}, {Rodriguez-Gomez}, {Sijacki}  \& {Springel}}{{Genel}
  et~al.}{2015}]{Genel2015}
{Genel} S.,  {Fall} S.~M.,  {Hernquist} L.,  {Vogelsberger} M.,  {Snyder}
  G.~F.,  {Rodriguez-Gomez} V.,  {Sijacki} D.,   {Springel} V.,  2015, \mn@doi
  [\apjl] {10.1088/2041-8205/804/2/L40}, \href
  {https://ui.adsabs.harvard.edu/abs/2015ApJ...804L..40G} {804, L40}

\bibitem[\protect\citeauthoryear{{Grieb}, {S{\'a}nchez}, {Salazar-Albornoz}  \&
  {Dalla Vecchia}}{{Grieb} et~al.}{2016}]{Grieb2016}
{Grieb} J.~N.,  {S{\'a}nchez} A.~G.,  {Salazar-Albornoz} S.,   {Dalla Vecchia}
  C.,  2016, \mn@doi [\mnras] {10.1093/mnras/stw065}, \href
  {https://ui.adsabs.harvard.edu/abs/2016MNRAS.457.1577G} {457, 1577}

\bibitem[\protect\citeauthoryear{{Guo} et~al.,}{{Guo} et~al.}{2015}]{Guo2015}
{Guo} H.,  et~al., 2015, \mn@doi [\mnras] {10.1093/mnras/stv1966}, \href
  {https://ui.adsabs.harvard.edu/abs/2015MNRAS.453.4368G} {453, 4368}

\bibitem[\protect\citeauthoryear{{Huang} et~al.,}{{Huang}
  et~al.}{2019}]{Huang2019}
{Huang} S.,  et~al., 2019, \mn@doi [\mnras] {10.1093/mnras/stz057}, \href
  {https://ui.adsabs.harvard.edu/abs/2019MNRAS.484.2021H} {484, 2021}

\bibitem[\protect\citeauthoryear{{Huchra} \& {Geller}}{{Huchra} \&
  {Geller}}{1982}]{Huchra1982}
{Huchra} J.~P.,  {Geller} M.~J.,  1982, \mn@doi [\apj] {10.1086/160000}, \href
  {https://ui.adsabs.harvard.edu/abs/1982ApJ...257..423H} {257, 423}

\bibitem[\protect\citeauthoryear{{Kauffmann}, {White}, {Heckman}, {M{\'e}nard},
  {Brinchmann}, {Charlot}, {Tremonti}  \& {Brinkmann}}{{Kauffmann}
  et~al.}{2004}]{Kauffmann2004}
{Kauffmann} G.,  {White} S. D.~M.,  {Heckman} T.~M.,  {M{\'e}nard} B.,
  {Brinchmann} J.,  {Charlot} S.,  {Tremonti} C.,   {Brinkmann} J.,  2004,
  \mn@doi [\mnras] {10.1111/j.1365-2966.2004.08117.x}, \href
  {https://ui.adsabs.harvard.edu/abs/2004MNRAS.353..713K} {353, 713}

\bibitem[\protect\citeauthoryear{{Kennicutt}}{{Kennicutt}}{1998}]{Kennicutt1998}
{Kennicutt} Robert~C. J.,  1998, \mn@doi [\araa]
  {10.1146/annurev.astro.36.1.189}, \href
  {https://ui.adsabs.harvard.edu/abs/1998ARA&A..36..189K} {36, 189}

\bibitem[\protect\citeauthoryear{{Kennicutt} \& {Evans}}{{Kennicutt} \&
  {Evans}}{2012}]{Kennicutt2012}
{Kennicutt} R.~C.,  {Evans} N.~J.,  2012, \mn@doi [\araa]
  {10.1146/annurev-astro-081811-125610}, \href
  {https://ui.adsabs.harvard.edu/abs/2012ARA&A..50..531K} {50, 531}

\bibitem[\protect\citeauthoryear{{Kim}, {Baugh}, {Cole}, {Frenk}  \&
  {Benson}}{{Kim} et~al.}{2009}]{Kim2009}
{Kim} H.-S.,  {Baugh} C.~M.,  {Cole} S.,  {Frenk} C.~S.,   {Benson} A.~J.,
  2009, \mn@doi [\mnras] {10.1111/j.1365-2966.2009.15560.x}, \href
  {https://ui.adsabs.harvard.edu/abs/2009MNRAS.400.1527K} {400, 1527}

\bibitem[\protect\citeauthoryear{{Kraljic} et~al.,}{{Kraljic}
  et~al.}{2018}]{Kraljic2018}
{Kraljic} K.,  et~al., 2018, \mn@doi [\mnras] {10.1093/mnras/stx2638}, \href
  {https://ui.adsabs.harvard.edu/abs/2018MNRAS.474..547K} {474, 547}

\bibitem[\protect\citeauthoryear{{Kravtsov}, {Berlind}, {Wechsler}, {Klypin},
  {Gottl{\"o}ber}, {Allgood}  \& {Primack}}{{Kravtsov}
  et~al.}{2004}]{Kravtsov2004}
{Kravtsov} A.~V.,  {Berlind} A.~A.,  {Wechsler} R.~H.,  {Klypin} A.~A.,
  {Gottl{\"o}ber} S.,  {Allgood} B.,   {Primack} J.~R.,  2004, \mn@doi [\apj]
  {10.1086/420959}, \href
  {https://ui.adsabs.harvard.edu/abs/2004ApJ...609...35K} {609, 35}

\bibitem[\protect\citeauthoryear{{Laigle} et~al.,}{{Laigle}
  et~al.}{2018}]{Laigle2018}
{Laigle} C.,  et~al., 2018, \mn@doi [\mnras] {10.1093/mnras/stx3055}, \href
  {https://ui.adsabs.harvard.edu/abs/2018MNRAS.474.5437L} {474, 5437}

\bibitem[\protect\citeauthoryear{{Lee} \& {Moon}}{{Lee} \&
  {Moon}}{2023}]{Lee2023}
{Lee} J.,  {Moon} J.-S.,  2023, \mn@doi [arXiv e-prints]
  {10.48550/arXiv.2305.04409}, \href
  {https://ui.adsabs.harvard.edu/abs/2023arXiv230504409L} {p. arXiv:2305.04409}

\bibitem[\protect\citeauthoryear{{Lietzen}, {Tempel}, {Hein{\"a}m{\"a}ki},
  {Nurmi}, {Einasto}  \& {Saar}}{{Lietzen} et~al.}{2012}]{Lietzen2012}
{Lietzen} H.,  {Tempel} E.,  {Hein{\"a}m{\"a}ki} P.,  {Nurmi} P.,  {Einasto}
  M.,   {Saar} E.,  2012, \mn@doi [\aap] {10.1051/0004-6361/201219353}, \href
  {https://ui.adsabs.harvard.edu/abs/2012A&A...545A.104L} {545, A104}

\bibitem[\protect\citeauthoryear{{Lifshitz}}{{Lifshitz}}{1946}]{Lifshitz46}
{Lifshitz} E.~M.,  1946, Zhurnal Eksperimentalnoi i Teoreticheskoi Fiziki,
  \href {https://ui.adsabs.harvard.edu/abs/1946ZhETF..16..587L} {16, 587}

\bibitem[\protect\citeauthoryear{{Luparello}, {Lares}, {Lambas}  \&
  {Padilla}}{{Luparello} et~al.}{2011}]{Luparello2011}
{Luparello} H.,  {Lares} M.,  {Lambas} D.~G.,   {Padilla} N.,  2011, \mn@doi
  [\mnras] {10.1111/j.1365-2966.2011.18794.x}, \href
  {https://ui.adsabs.harvard.edu/abs/2011MNRAS.415..964L} {415, 964}

\bibitem[\protect\citeauthoryear{{Malavasi} et~al.,}{{Malavasi}
  et~al.}{2017}]{Malavasi2017}
{Malavasi} N.,  et~al., 2017, \mn@doi [\mnras] {10.1093/mnras/stw2864}, \href
  {https://ui.adsabs.harvard.edu/abs/2017MNRAS.465.3817M} {465, 3817}

\bibitem[\protect\citeauthoryear{Mansfield \& Kravtsov}{Mansfield \&
  Kravtsov}{2020}]{Mansfield2020}
Mansfield P.,  Kravtsov A.~V.,  2020, Monthly Notices of the Royal Astronomical
  Society, 493, 4763

\bibitem[\protect\citeauthoryear{{Mao}, {Zentner}  \& {Wechsler}}{{Mao}
  et~al.}{2018}]{Mao2018}
{Mao} Y.-Y.,  {Zentner} A.~R.,   {Wechsler} R.~H.,  2018, \mn@doi [\mnras]
  {10.1093/mnras/stx3111}, \href
  {https://ui.adsabs.harvard.edu/abs/2018MNRAS.474.5143M} {474, 5143}

\bibitem[\protect\citeauthoryear{{Marinacci} et~al.,}{{Marinacci}
  et~al.}{2018}]{Marinacci2018}
{Marinacci} F.,  et~al., 2018, \mn@doi [\mnras] {10.1093/mnras/sty2206}, \href
  {https://ui.adsabs.harvard.edu/abs/2018MNRAS.480.5113M} {480, 5113}

\bibitem[\protect\citeauthoryear{{Mart{\'\i}nez}, {Coenda}  \&
  {Muriel}}{{Mart{\'\i}nez} et~al.}{2008}]{Martinez2008}
{Mart{\'\i}nez} H.~J.,  {Coenda} V.,   {Muriel} H.,  2008, \mn@doi [\mnras]
  {10.1111/j.1365-2966.2008.13929.x}, \href
  {https://ui.adsabs.harvard.edu/abs/2008MNRAS.391..585M} {391, 585}

\bibitem[\protect\citeauthoryear{{Moutard}, {Sawicki}, {Arnouts}, {Golob},
  {Malavasi}, {Adami}, {Coupon}  \& {Ilbert}}{{Moutard}
  et~al.}{2018}]{Moutard2018}
{Moutard} T.,  {Sawicki} M.,  {Arnouts} S.,  {Golob} A.,  {Malavasi} N.,
  {Adami} C.,  {Coupon} J.,   {Ilbert} O.,  2018, \mn@doi [\mnras]
  {10.1093/mnras/sty1543}, \href
  {https://ui.adsabs.harvard.edu/abs/2018MNRAS.479.2147M} {479, 2147}

\bibitem[\protect\citeauthoryear{Musso, Cadiou, Pichon, Codis, Kraljic  \&
  Dubois}{Musso et~al.}{2018}]{Musso2018}
Musso M.,  Cadiou C.,  Pichon C.,  Codis S.,  Kraljic K.,   Dubois Y.,  2018,
  Monthly Notices of the Royal Astronomical Society, 476, 4877

\bibitem[\protect\citeauthoryear{{Naiman} et~al.,}{{Naiman}
  et~al.}{2018}]{Naiman2018}
{Naiman} J.~P.,  et~al., 2018, \mn@doi [\mnras] {10.1093/mnras/sty618}, \href
  {https://ui.adsabs.harvard.edu/abs/2018MNRAS.477.1206N} {477, 1206}

\bibitem[\protect\citeauthoryear{{Nelson} et~al.,}{{Nelson}
  et~al.}{2018}]{Nelson2018}
{Nelson} D.,  et~al., 2018, \mn@doi [\mnras] {10.1093/mnras/stx3040}, \href
  {https://ui.adsabs.harvard.edu/abs/2018MNRAS.475..624N} {475, 624}

\bibitem[\protect\citeauthoryear{{Osato} \& {Okumura}}{{Osato} \&
  {Okumura}}{2023}]{Osato2023}
{Osato} K.,  {Okumura} T.,  2023, \mn@doi [\mnras] {10.1093/mnras/stac3582},
  \href {https://ui.adsabs.harvard.edu/abs/2023MNRAS.519.1771O} {519, 1771}

\bibitem[\protect\citeauthoryear{{Peacock} \& {Smith}}{{Peacock} \&
  {Smith}}{2000}]{Peacock2000}
{Peacock} J.~A.,  {Smith} R.~E.,  2000, \mn@doi [\mnras]
  {10.1046/j.1365-8711.2000.03779.x}, \href
  {https://ui.adsabs.harvard.edu/abs/2000MNRAS.318.1144P} {318, 1144}

\bibitem[\protect\citeauthoryear{{Peebles}}{{Peebles}}{1980}]{Peebles1980}
{Peebles} P.~J.~E.,  1980, {The large-scale structure of the universe}

\bibitem[\protect\citeauthoryear{{Peng} et~al.,}{{Peng}
  et~al.}{2010}]{Peng2010}
{Peng} Y.-j.,  et~al., 2010, \mn@doi [\apj] {10.1088/0004-637X/721/1/193},
  \href {https://ui.adsabs.harvard.edu/abs/2010ApJ...721..193P} {721, 193}

\bibitem[\protect\citeauthoryear{{Pillepich} et~al.,}{{Pillepich}
  et~al.}{2018a}]{Pillepich2018_a}
{Pillepich} A.,  et~al., 2018a, \mn@doi [\mnras] {10.1093/mnras/stx2656}, \href
  {https://ui.adsabs.harvard.edu/abs/2018MNRAS.473.4077P} {473, 4077}

\bibitem[\protect\citeauthoryear{{Pillepich} et~al.,}{{Pillepich}
  et~al.}{2018b}]{Pillepich2018_b}
{Pillepich} A.,  et~al., 2018b, \mn@doi [\mnras] {10.1093/mnras/stx3112}, \href
  {https://ui.adsabs.harvard.edu/abs/2018MNRAS.475..648P} {475, 648}

\bibitem[\protect\citeauthoryear{{Pillepich} et~al.,}{{Pillepich}
  et~al.}{2019}]{Pillepich2019}
{Pillepich} A.,  et~al., 2019, \mn@doi [\mnras] {10.1093/mnras/stz2338}, \href
  {https://ui.adsabs.harvard.edu/abs/2019MNRAS.490.3196P} {490, 3196}

\bibitem[\protect\citeauthoryear{{Planck Collaboration} et~al.,}{{Planck
  Collaboration} et~al.}{2016}]{planckresults}
{Planck Collaboration} et~al., 2016, \mn@doi [A\&A]
  {10.1051/0004-6361/201525830}, 594, A13

\bibitem[\protect\citeauthoryear{Quenouille}{Quenouille}{1949}]{Quenouille1949}
Quenouille M.~H.,  1949, \mn@doi [The Annals of Mathematical Statistics]
  {10.1214/aoms/1177729989}, 20, 355

\bibitem[\protect\citeauthoryear{Ramakrishnan, Paranjape, Hahn  \&
  Sheth}{Ramakrishnan et~al.}{2019}]{Ramakrishnan2019}
Ramakrishnan S.,  Paranjape A.,  Hahn O.,   Sheth R.~K.,  2019, Monthly Notices
  of the Royal Astronomical Society, 489, 2977

\bibitem[\protect\citeauthoryear{Rodriguez \& Merch{\'a}n}{Rodriguez \&
  Merch{\'a}n}{2020}]{Rodriguez2020}
Rodriguez F.,  Merch{\'a}n M.,  2020, Astronomy \& Astrophysics, 636, A61

\bibitem[\protect\citeauthoryear{{Rodriguez}, {Merch{\'a}n}  \&
  {Sgr{\'o}}}{{Rodriguez} et~al.}{2015}]{Rodriguez2015}
{Rodriguez} F.,  {Merch{\'a}n} M.,   {Sgr{\'o}} M.~A.,  2015, \mn@doi [\aap]
  {10.1051/0004-6361/201525798}, \href
  {https://ui.adsabs.harvard.edu/abs/2015A&A...580A..86R} {580, A86}

\bibitem[\protect\citeauthoryear{{Ross}, {Percival}  \& {Brunner}}{{Ross}
  et~al.}{2010}]{Ross2010}
{Ross} A.~J.,  {Percival} W.~J.,   {Brunner} R.~J.,  2010, \mn@doi [\mnras]
  {10.1111/j.1365-2966.2010.16914.x}, \href
  {https://ui.adsabs.harvard.edu/abs/2010MNRAS.407..420R} {407, 420}

\bibitem[\protect\citeauthoryear{{Salcedo}, {Maller}, {Berlind}, {Sinha},
  {McBride}, {Behroozi}, {Wechsler}  \& {Weinberg}}{{Salcedo}
  et~al.}{2018}]{Salcedo2018}
{Salcedo} A.~N.,  {Maller} A.~H.,  {Berlind} A.~A.,  {Sinha} M.,  {McBride}
  C.~K.,  {Behroozi} P.~S.,  {Wechsler} R.~H.,   {Weinberg} D.~H.,  2018,
  \mn@doi [\mnras] {10.1093/mnras/sty109}, \href
  {https://ui.adsabs.harvard.edu/abs/2018MNRAS.475.4411S} {475, 4411}

\bibitem[\protect\citeauthoryear{{Schaap} \& {van de Weygaert}}{{Schaap} \&
  {van de Weygaert}}{2000}]{Schaap2000}
{Schaap} W.~E.,  {van de Weygaert} R.,  2000, \mn@doi [\aap]
  {10.48550/arXiv.astro-ph/0011007}, \href
  {https://ui.adsabs.harvard.edu/abs/2000A&A...363L..29S} {363, L29}

\bibitem[\protect\citeauthoryear{{Schawinski} et~al.,}{{Schawinski}
  et~al.}{2014}]{Schawinski2014}
{Schawinski} K.,  et~al., 2014, \mn@doi [\mnras] {10.1093/mnras/stu327}, \href
  {https://ui.adsabs.harvard.edu/abs/2014MNRAS.440..889S} {440, 889}

\bibitem[\protect\citeauthoryear{{Shandarin} \& {Zeldovich}}{{Shandarin} \&
  {Zeldovich}}{1989}]{shandarin89}
{Shandarin} S.~F.,  {Zeldovich} Y.~B.,  1989, \mn@doi [Reviews of Modern
  Physics] {10.1103/RevModPhys.61.185}, \href
  {https://ui.adsabs.harvard.edu/abs/1989RvMP...61..185S} {61, 185}

\bibitem[\protect\citeauthoryear{{Sousbie}}{{Sousbie}}{2011}]{Sousbie2011_a}
{Sousbie} T.,  2011, \mn@doi [\mnras] {10.1111/j.1365-2966.2011.18394.x}, \href
  {https://ui.adsabs.harvard.edu/abs/2011MNRAS.414..350S} {414, 350}

\bibitem[\protect\citeauthoryear{{Sousbie}}{{Sousbie}}{2013}]{Sousbie2013}
{Sousbie} T.,  2013, {DisPerSE: Discrete Persistent Structures Extractor},
  Astrophysics Source Code Library, record ascl:1302.015 (\mn@eprint {ascl}
  {1302.015})

\bibitem[\protect\citeauthoryear{{Sousbie}, {Pichon}  \& {Kawahara}}{{Sousbie}
  et~al.}{2011}]{Sousbie2011_b}
{Sousbie} T.,  {Pichon} C.,   {Kawahara} H.,  2011, \mn@doi [\mnras]
  {10.1111/j.1365-2966.2011.18395.x}, \href
  {https://ui.adsabs.harvard.edu/abs/2011MNRAS.414..384S} {414, 384}

\bibitem[\protect\citeauthoryear{{Springel}}{{Springel}}{2010}]{Springel2010}
{Springel} V.,  2010, \mn@doi [\mnras] {10.1111/j.1365-2966.2009.15715.x},
  \href {https://ui.adsabs.harvard.edu/abs/2010MNRAS.401..791S} {401, 791}

\bibitem[\protect\citeauthoryear{{Springel}, {White}, {Tormen}  \&
  {Kauffmann}}{{Springel} et~al.}{2001}]{Springel2001}
{Springel} V.,  {White} S. D.~M.,  {Tormen} G.,   {Kauffmann} G.,  2001,
  \mn@doi [\mnras] {10.1046/j.1365-8711.2001.04912.x}, \href
  {https://ui.adsabs.harvard.edu/abs/2001MNRAS.328..726S} {328, 726}

\bibitem[\protect\citeauthoryear{{Springel} et~al.,}{{Springel}
  et~al.}{2018}]{Springel2018}
{Springel} V.,  et~al., 2018, \mn@doi [\mnras] {10.1093/mnras/stx3304}, \href
  {https://ui.adsabs.harvard.edu/abs/2018MNRAS.475..676S} {475, 676}

\bibitem[\protect\citeauthoryear{{Tempel} \& {Libeskind}}{{Tempel} \&
  {Libeskind}}{2013}]{Tempel2013_2}
{Tempel} E.,  {Libeskind} N.~I.,  2013, \mn@doi [\apjl]
  {10.1088/2041-8205/775/2/L42}, \href
  {https://ui.adsabs.harvard.edu/abs/2013ApJ...775L..42T} {775, L42}

\bibitem[\protect\citeauthoryear{{Tempel}, {Stoica}  \& {Saar}}{{Tempel}
  et~al.}{2013}]{Tempel2013}
{Tempel} E.,  {Stoica} R.~S.,   {Saar} E.,  2013, \mn@doi [\mnras]
  {10.1093/mnras/sts162}, \href
  {https://ui.adsabs.harvard.edu/abs/2013MNRAS.428.1827T} {428, 1827}

\bibitem[\protect\citeauthoryear{{Tinker}, {Cao}, {Alpaslan}, {DeRose}, {Mao}
  \& {Wechsler}}{{Tinker} et~al.}{2021}]{Tinker2021}
{Tinker} J.~L.,  {Cao} J.,  {Alpaslan} M.,  {DeRose} J.,  {Mao} Y.-Y.,
  {Wechsler} R.~H.,  2021, \mn@doi [\mnras] {10.1093/mnras/stab1576}, \href
  {https://ui.adsabs.harvard.edu/abs/2021MNRAS.505.5370T} {505, 5370}

\bibitem[\protect\citeauthoryear{{Trujillo}, {Carretero}  \&
  {Patiri}}{{Trujillo} et~al.}{2006}]{Trujillo2006}
{Trujillo} I.,  {Carretero} C.,   {Patiri} S.~G.,  2006, \mn@doi [\apjl]
  {10.1086/503548}, \href
  {https://ui.adsabs.harvard.edu/abs/2006ApJ...640L.111T} {640, L111}

\bibitem[\protect\citeauthoryear{Tukey}{Tukey}{1958}]{Tukey1958}
Tukey J.~W.,  1958, \mn@doi [The Annals of Mathematical Statistics]
  {10.1214/aoms/1177706647}, 29, 614

\bibitem[\protect\citeauthoryear{{Vogelsberger} et~al.,}{{Vogelsberger}
  et~al.}{2014a}]{Vogelsberger2014_a}
{Vogelsberger} M.,  et~al., 2014a, \mn@doi [\mnras] {10.1093/mnras/stu1536},
  \href {https://ui.adsabs.harvard.edu/abs/2014MNRAS.444.1518V} {444, 1518}

\bibitem[\protect\citeauthoryear{{Vogelsberger} et~al.,}{{Vogelsberger}
  et~al.}{2014b}]{Vogelsberger2014_b}
{Vogelsberger} M.,  et~al., 2014b, \mn@doi [\nat] {10.1038/nature13316}, \href
  {https://ui.adsabs.harvard.edu/abs/2014Natur.509..177V} {509, 177}

\bibitem[\protect\citeauthoryear{{Vulcani}, {Poggianti}, {Fritz}, {Fasano},
  {Moretti}, {Calvi}  \& {Paccagnella}}{{Vulcani} et~al.}{2015}]{Vulcani2015}
{Vulcani} B.,  {Poggianti} B.~M.,  {Fritz} J.,  {Fasano} G.,  {Moretti} A.,
  {Calvi} R.,   {Paccagnella} A.,  2015, \mn@doi [\apj]
  {10.1088/0004-637X/798/1/52}, \href
  {https://ui.adsabs.harvard.edu/abs/2015ApJ...798...52V} {798, 52}

\bibitem[\protect\citeauthoryear{{Wang}, {Libeskind}, {Tempel}, {Pawlowski},
  {Kang}  \& {Guo}}{{Wang} et~al.}{2020}]{Wang2020}
{Wang} P.,  {Libeskind} N.~I.,  {Tempel} E.,  {Pawlowski} M.~S.,  {Kang} X.,
  {Guo} Q.,  2020, \mn@doi [\apj] {10.3847/1538-4357/aba6ea}, \href
  {https://ui.adsabs.harvard.edu/abs/2020ApJ...900..129W} {900, 129}

\bibitem[\protect\citeauthoryear{{Weinmann}, {van den Bosch}, {Yang}  \&
  {Mo}}{{Weinmann} et~al.}{2006}]{Weinmann2006}
{Weinmann} S.~M.,  {van den Bosch} F.~C.,  {Yang} X.,   {Mo} H.~J.,  2006,
  \mn@doi [\mnras] {10.1111/j.1365-2966.2005.09865.x}, \href
  {https://ui.adsabs.harvard.edu/abs/2006MNRAS.366....2W} {366, 2}

\bibitem[\protect\citeauthoryear{White}{White}{1994}]{White1994}
White S. D.~M.,  1994, Formation and Evolution of Galaxies: Les Houches
  Lectures (\mn@eprint {arXiv} {astro-ph/9410043})

\bibitem[\protect\citeauthoryear{{White} \& {Rees}}{{White} \&
  {Rees}}{1978}]{WhiteRees78}
{White} S.~D.~M.,  {Rees} M.~J.,  1978, \mn@doi [\mnras]
  {10.1093/mnras/183.3.341}, \href
  {https://ui.adsabs.harvard.edu/abs/1978MNRAS.183..341W} {183, 341}

\bibitem[\protect\citeauthoryear{{Yuan}, {Hadzhiyska}, {Bose}  \&
  {Eisenstein}}{{Yuan} et~al.}{2022}]{Yuan2022}
{Yuan} S.,  {Hadzhiyska} B.,  {Bose} S.,   {Eisenstein} D.~J.,  2022, \mn@doi
  [\mnras] {10.1093/mnras/stac830}, \href
  {https://ui.adsabs.harvard.edu/abs/2022MNRAS.512.5793Y} {512, 5793}

\bibitem[\protect\citeauthoryear{{Zehavi} et~al.,}{{Zehavi}
  et~al.}{2011}]{Zehavi2011}
{Zehavi} I.,  et~al., 2011, \mn@doi [\apj] {10.1088/0004-637X/736/1/59}, \href
  {https://ui.adsabs.harvard.edu/abs/2011ApJ...736...59Z} {736, 59}

\bibitem[\protect\citeauthoryear{{Zehavi}, {Contreras}, {Padilla}, {Smith},
  {Baugh}  \& {Norberg}}{{Zehavi} et~al.}{2018}]{Zehavi2018}
{Zehavi} I.,  {Contreras} S.,  {Padilla} N.,  {Smith} N.~J.,  {Baugh} C.~M.,
  {Norberg} P.,  2018, \mn@doi [\apj] {10.3847/1538-4357/aaa54a}, \href
  {https://ui.adsabs.harvard.edu/abs/2018ApJ...853...84Z} {853, 84}

\bibitem[\protect\citeauthoryear{{Zel'dovich}}{{Zel'dovich}}{1970}]{Zel'dovich1970}
{Zel'dovich} Y.~B.,  1970, \aap, \href
  {https://ui.adsabs.harvard.edu/abs/1970A&A.....5...84Z} {5, 84}

\bibitem[\protect\citeauthoryear{{Zhang}, {Yang}, {Wang}, {Wang}, {Luo}, {Mo}
  \& {van den Bosch}}{{Zhang} et~al.}{2015}]{Zhang2015}
{Zhang} Y.,  {Yang} X.,  {Wang} H.,  {Wang} L.,  {Luo} W.,  {Mo} H.~J.,   {van
  den Bosch} F.~C.,  2015, \mn@doi [\apj] {10.1088/0004-637X/798/1/17}, \href
  {https://ui.adsabs.harvard.edu/abs/2015ApJ...798...17Z} {798, 17}

\bibitem[\protect\citeauthoryear{{Zheng} \& {Weinberg}}{{Zheng} \&
  {Weinberg}}{2007}]{Zheng2007}
{Zheng} Z.,  {Weinberg} D.~H.,  2007, \mn@doi [\apj] {10.1086/512151}, \href
  {https://ui.adsabs.harvard.edu/abs/2007ApJ...659....1Z} {659, 1}

\bibitem[\protect\citeauthoryear{{Zheng} et~al.,}{{Zheng}
  et~al.}{2005}]{Zheng2005}
{Zheng} Z.,  et~al., 2005, \mn@doi [\apj] {10.1086/466510}, \href
  {https://ui.adsabs.harvard.edu/abs/2005ApJ...633..791Z} {633, 791}

\bibitem[\protect\citeauthoryear{{de Lapparent}, {Geller}  \& {Huchra}}{{de
  Lapparent} et~al.}{1986}]{deLapparent1986}
{de Lapparent} V.,  {Geller} M.~J.,   {Huchra} J.~P.,  1986, \mn@doi [\apjl]
  {10.1086/184625}, \href
  {https://ui.adsabs.harvard.edu/abs/1986ApJ...302L...1D} {302, L1}

\bibitem[\protect\citeauthoryear{{van de Weygaert} \& {Schaap}}{{van de
  Weygaert} \& {Schaap}}{2009}]{vandeWeygaert2009}
{van de Weygaert} R.,  {Schaap} W.,  2009, in {Mart{\'\i}nez} V.~J.,  {Saar}
  E.,  {Mart{\'\i}nez-Gonz{\'a}lez} E.,   {Pons-Border{\'\i}a} M.~J.,  eds, ,
  Vol.~665, Data Analysis in Cosmology.
pp 291--413, \mn@doi{10.1007/978-3-540-44767-2_11}

\bibitem[\protect\citeauthoryear{{van den Bosch}, {Yang}  \& {Mo}}{{van den
  Bosch} et~al.}{2003}]{vandenBosch2003}
{van den Bosch} F.~C.,  {Yang} X.,   {Mo} H.~J.,  2003, \mn@doi [\mnras]
  {10.1046/j.1365-8711.2003.06335.x}, \href
  {https://ui.adsabs.harvard.edu/abs/2003MNRAS.340..771V} {340, 771}

\makeatother
\end{thebibliography}




\appendix

\section{HOD measurements regarding to the SFR parameter}

The evolutionary history of galaxies can be inferred from the SFR parameter, which is related to the amount of gas and the efficiency of star formation \citep{Kennicutt1998}. 
Several processes at scales between parsecs and Megaparsecs affect star formation \citep{Kennicutt2012}, such as environment and galaxy mass \citep{Kauffmann2004,Peng2010, Moutard2018}.
With this in mind, in order to complement the HODs measurements with respect to the galaxy properties of Section \ref{sec:analysis}, we explore the SFR parameter.
Following \cite{Pillepich2019}, we calculate recursively star forming main sequence (SFMS) ridge line in the SFR vs. stellar mass plane. Then, we define star-forming and quiescent galaxies as those above and below the SFMS line, respectively.

\subsection{Filaments}
\begin{figure}
\centering
    {
        \includegraphics[width=0.47\textwidth]{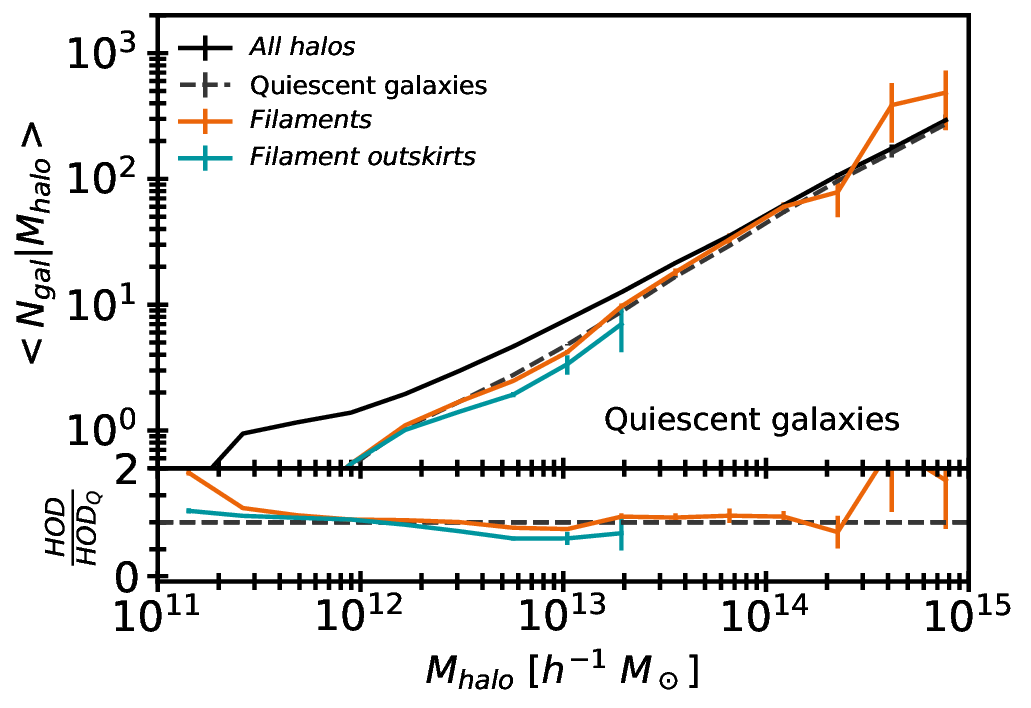}
    }
    {
        \includegraphics[width=0.47\textwidth]{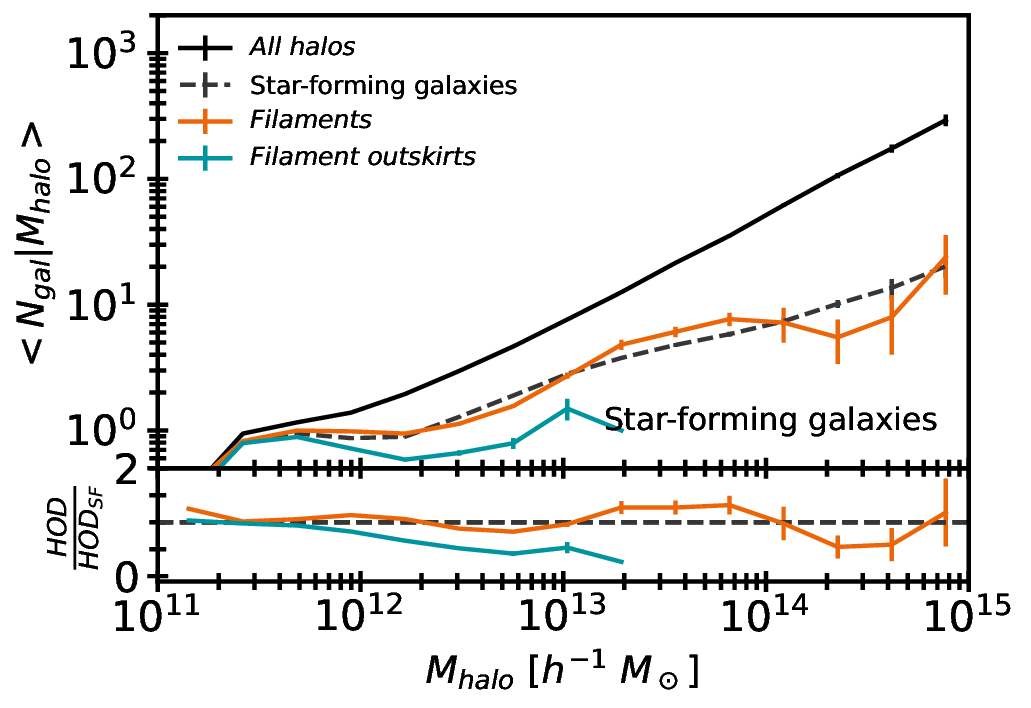}
    }
    \caption{Panels shows the HODs measured with respect to the subhalo star formation rate parameter (SFR) for the total (grey), filaments (orange) and filament outskirts (light-blue) samples for subhalos with $\text{M}_\text{r} - 5\text{log}_{10}(\text{h}) \leq -17$. The dashed lines represent the HODs for the total of quiescent (upper) and star-forming (lower) subhalos. Lower sub-panels show the ratio between the filaments and filament outskirts samples with the total of quiescent and star-forming subhalos, respectively.}
    \label{fig:hod_fil_sfr}
\end{figure}

The HODs of quiescent and star-forming galaxies 
are shown in Fig. \ref{fig:hod_fil_sfr}.
%
Looking at the quiescent galaxies, we observe that the filaments sample lies on the total distribution of inactive galaxies. For high-mass halos, the behaviour shows scattering caused by the low number of objects in these mass bins.
The trend in the filament outskirts is comparable to that of the sample within the error bars. 
This implies that inactive galaxies do not display a preference for any of these environments.
On the other hand, for low-mass halos, we observe that the star-forming galaxies in the filaments sample have a HOD similar to the total of the star-forming galaxies, while for high-mass halos the behaviours show a sinusoidal shape with respect to this distribution.
Besides, halos in the filament outskirts are practically devoid of star-forming galaxies.
The results discovered at the periphery of the filament resemble those found in the void regions by \cite{Alfaro2020}.

\subsection{Nodes}
\begin{figure}
\centering
    {
        \includegraphics[width=0.47\textwidth]{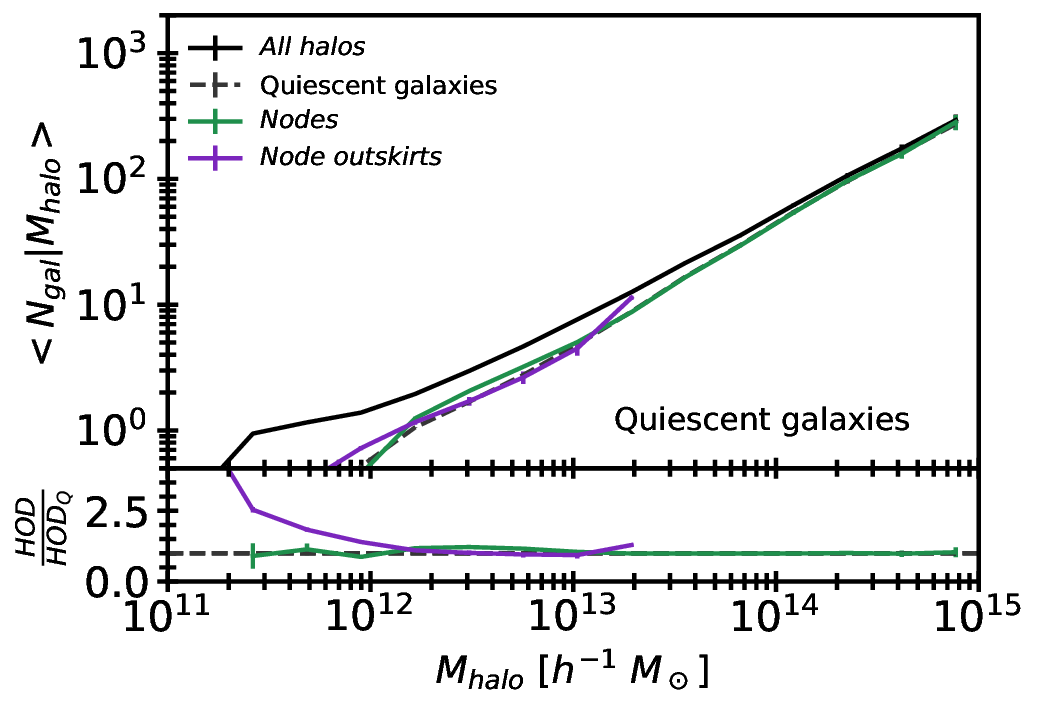}
    }
    {
        \includegraphics[width=0.47\textwidth]{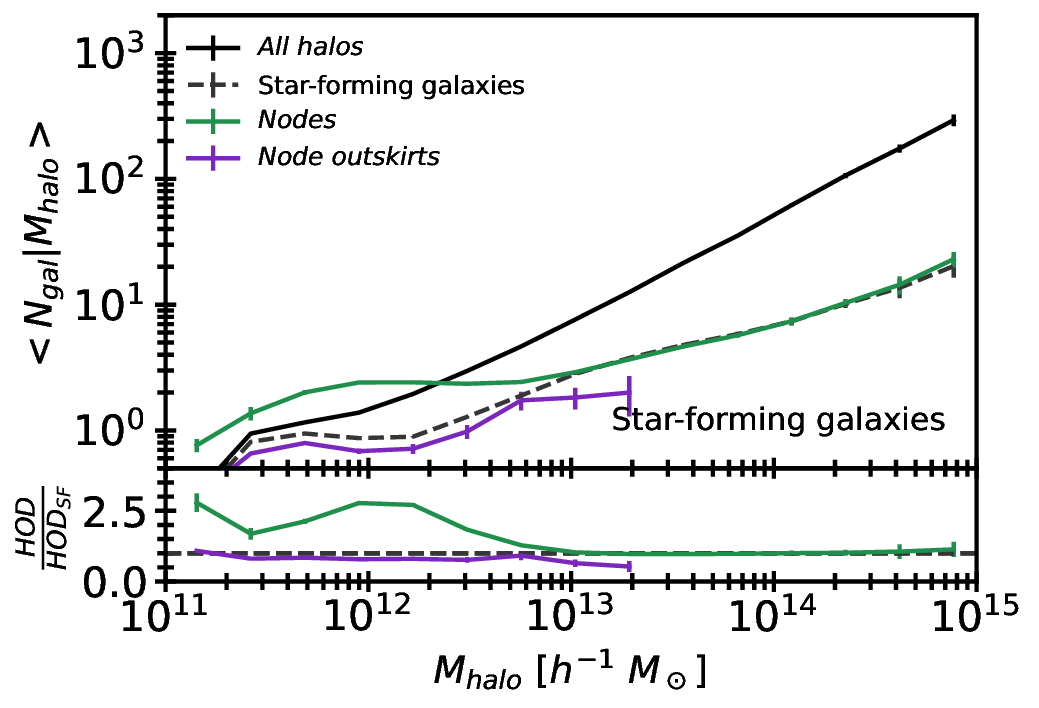}
    }
    \caption{Panel shows the measured HOD with respect to the SFR parameter for the total (grey), nodes (green) and node outskirts (purple) samples for subhalos with $\text{M}_\text{r} - 5\text{log}_{10}(\text{h}) \leq -17$. The dashed lines represent the HODs for the total of quiescent (upper) and star-forming (lower) subhalos. Lower sub-panels show the ratio between the nodes and node outskirts samples with the total of quiescent and star-forming subhalos, respectively.}
    \label{fig:hod_nod_sfr}
\end{figure}

In addition, in Fig. \ref{fig:hod_nod_sfr} we show the HODs taking into account the SFR parameter. 
%
In the upper panel we observe the HODs for quiescent galaxies. 
The distributions of both samples lie on the HOD for total inactive subhalos.
This would mean that the occupancy by this type of galaxy is less dependent on specific environmental conditions.
Finally, the lower panel shows the distributions for star-forming galaxies.
Low-mass halos in nodes have an excess of star-forming galaxies over the total active galaxies, which is even higher than the total feature, then for massive halos the HOD decreases and is placed on the trend of the total star-forming galaxies.
This may be related to the fact that the hot gas fraction decreases at low halo masses \citep{Huang2019}, favouring star formation in these halos.
On the contrary, halos in the regions surrounding the nodes have a lower distribution than the total active galaxy distribution for the whole range of masses. 

\bsp	
\label{lastpage}
\end{document}